\documentclass[iop]{emulateapj}

\usepackage{amsmath}
\usepackage{amssymb}
\usepackage{graphicx}
\usepackage{graphics}
\usepackage{subfigure}
\usepackage{threeparttable}
\usepackage{astron}
\usepackage{natbib}
\usepackage{array}
\usepackage[perpage]{footmisc}
\usepackage{rotating}
\usepackage{textcomp}
\usepackage{color}
\usepackage{listings}
\usepackage{wrapfig}
\usepackage{sidecap}
\usepackage{float}
\usepackage{subfloat}
\usepackage{hyperref}

\begin{document}
\fontsize{10}{10}\selectfont

\bibliographystyle{apj}

\slugcomment{Accepted for publication in ApJ; \today}

\shorttitle{Resonant Clumping}
\shortauthors{Molloy et al.}

\title{Resonant Clumping and Substructure in Galactic Discs}
\author{Matthew Molloy}
\affil{Kavli Institute for Astronomy \& Astrophysics, Peking University, \\ 
Yi He Yuan Lu 5, Hai Dian Qu, Beijing 100871, China}
\author{Martin C. Smith\altaffilmark{$\star$} and Juntai Shen\altaffilmark{$\star$}}
\affil{Key Laboratory for Research in Galaxies and Cosmology, Shanghai Astronomical Observatory, \\ 
Chinese Academy of Sciences, 80 Nandan Road, Shanghai 200030, China}
\author{N. Wyn Evans}
\affil{Institute of Astronomy, Madingley Road, Cambridge, CB3 0HA, UK}

\altaffiltext{$\star$}{Corresponding Authors: matthewmolloy@gmail.com (MM), msmith@shao.ac.cn (MCS), jshen@shao.ac.cn (JS), nwe@ast.cam.ac.uk (NWE)}

\begin{abstract}
We describe a method to extract resonant orbits from N-body
simulations exploiting the fact that they close in frames rotating
with a constant pattern speed.  Our method is applied to the N-body
simulation of the Milky Way by \citet{Shen2010}.  This simulation
hosts a massive bar, which drives strong resonances and persistent
angular momentum exchange.  Resonant orbits are found throughout the
disc, both close to the bar and out to the very edges of the
disc.  Using Fourier spectrograms, we demonstrate that the bar is
driving kinematic substructure even in the very outer parts of the
disc. We identify two major orbit families in the outskirts of the
disc, one of which makes significant contributions to the kinematic landscape,
namely the $m:l$ = $3:-2$ family resonating with the bar.

A mechanism is described that produces bimodal distributions of
Galactocentric radial velocities at selected azimuths in the outer
disc. It occurs as a result of the temporal coherence of particles on
the $3:-2$ resonant orbits, which causes them to arrive simultaneously
at pericentre or apocentre.  This {\it resonant clumping}, due to the
in-phase motion of the particles through their epicycle, leads to both
inward and outward moving groups which belong to the same orbital
family and consequently produce bimodal radial velocity distributions.
This is a possible explanation of the bimodal velocity distributions
observed towards the Galactic anti-Centre by \citet{Liu2012a}.
Another consequence is that transient overdensities appear and
dissipate (in a symmetric fashion) resulting in a periodic pulsing of the disc's surface
density.
\end{abstract}

\section{Introduction}

Modelling the stellar disc in spiral galaxies often begins with the
assumption of axisymmetry.  For some purposes, this can be sufficient
and is often illuminating and powerful.  Real spiral galaxies however
are not so simple and exhibit a wealth of structure leading to strong
non-axisymmetries.  The Milky Way is host to numerous non-axisymmetric
features such as spiral arms and the central bar. It has been shown
that the majority of bright spirals are host to strong bars
\citep[e.g.,][]{Eskridge2000,Marinova2007}.  N-body simulations of
disc galaxies that start out in an axisymmetric configuration often
succumb to dynamical instabilities and self-consistently develop bars
and spirals
\citep[e.g.,][]{Sellwood1981,Debattista2006,Martinez-Valpuesta2006}.
These features can have have a significant effect on the orbital
structure in discs and so it is of great importance to understand
these mechanisms and have the right tools at our disposal to analyse
models that exhibit these features.

Orbits in non-axisymmetric systems periodically experience torques
introduced by rotating overdensities.  Long lived features allow time
for orbits to become resonantly coupled with periodic variations in
the torque as the disc evolves.  Naturally, this means that the
locations of these resonances are strongly linked to the pattern speed
of the rotating features and the rotation curve of the disc.  The
important locations are the resonant radii such as at corotation (CR)
and the location of the Lindblad resonances (LR) \citep[see, for
  example,][]{Weinberg1994,BT2}.  At these locations, the orbits
differ drastically from those in axisymmetric models.  Deviations from
axisymmetry, even in the disc centre, therefore cannot be disregarded
when trying to explain kinematic structure in the outskirts of the
disc.  While it has been shown in external galaxies and in simulations
that a central bar cannot extend beyond CR
\citep{Chirikov1979,Contopoulos1980,Sellwood1993}, it can have
observable effects and influence on the dynamics even in the outer
parts of the disc.

It is now beyond doubt that the Milky Way is host to a central bar
\citep[e.g.,][]{Blitz1991,Dwek1995}, yet its characteristic properties
are still the subject of debate
\citep[e.g.,][]{Dehnen1999,Gerhard2012}.  Resonant features induced by
the bar have been used to explain the nature of moving groups in the
Solar vicinity seen in the Hipparcos and Geneva-Copenhagen Survey data
\citep{Dehnen2000,Minchev2010}.  Departures from axisymmetry in the disc have
also been suggested as the reason for high values of the vertex deviation
in the Solar neighbourhood and also low values for the ratio of the
principal axes of the velocity-dispersion
tensor~\citep{Evans1993,DehnenBinney1998}.  Larger scale surveys such
as the RAVE \citep[RAdial Velocity Experiment;][]{Steinmetz2006} have
recently been used to map the kinematic landscape in the Solar
neighbourhood \citep[e.g.,][]{Siebert2011,Monari2014}, but the lack of
accurate parallaxes and proper motions prohibit an investigation into
full 6-D phase-space structures over significantly broad areas of the
disc.  The Sloan Digital Sky Survey \citep[SDSS;][]{Ahn2012} and in
particular the Sloan Extension for Galactic Understanding and
Exploration \citep[SEGUE;][]{Yanny2009} has been of immense value in
describing the properties of the Milky Way disc
\citep[e.g.,][]{Carollo2010,Bond2010,Bovy2012b,Bovy2012a,Smith2012},
in uncovering substructure in the Milky Way's stellar halo
\citep[e.g.,][]{Smith2009,Yanny2009a,deJong2010} and in discovering
new Milky Way satellites and characterising their tidal tails
\citep[e.g.,][]{Newberg2009,Belokurov2010,Koposov2010}.  More recently
the APOGEE \citep[Apache Point Observatory Galactic Evolution
  Experiment;][]{Eisenstein2011,Majewski2014} infrared (IR) survey has
probed the low latitude regions of the disc, heavily obscured in
optical surveys.  This large scale IR survey of the Galactic disc
(mostly bright giants and red clump (RC) stars) has allowed us to
identify kinematic features associated with the bar
\citep[e.g.,][]{2012Nidever,Vasquez2013}, but without accurate
distances on the largest scales their interpretation is still the
subject of debate \citep[see][]{Zhao-Yu2014}.  On smaller scales, but
with more accurate distances, the RC sample \citep{Bovy2014} has
yielded valuable insights into broad streaming motions present in the
disc \citep{Bovy2014arXiv} as well as motions associated with the
spiral arms \citep{Kawata2014} and the Galactic warp
\citep{Lopez-Corredoira2014arXiv}.  Beyond positions and velocities,
spectroscopic surveys also permit a view of the chemical landscape of
the disc
\citep{Randich2013,Rojas-Arriagada2014,Howes2014,Nidever2014}.
Obtaining accurate 6D phase space information along with abundances
allows us to probe the evolutionary history of the Milky Way.

Orbits in stellar discs are often described by oscillations in the
radial, azimuthal and vertical directions with respective frequencies,
$\kappa$, $\Omega$ and $\nu$.  Non-axisymmetric patterns are also
characterised by the frequency with which they rotate in the disc,
often called the pattern speed, $\Omega_{p}$.  A resonance occurs (in
the plane of the disk) when there is a commensurability between these
frequencies, namely
\begin{equation}
\label{eqn:resCond}
l \kappa=m(\Omega-\Omega_{\rm p}),
\end{equation}
or equivalently,
\begin{equation}
\label{eqn:resCond2}
lT'_{\phi}=mT_{R},
\end{equation}
where $l$ and $m$ are some integer values and $T_{R}$ and $T'_{\phi}$
are the epicyclic and the orbital period in the rotating frame
respectively.  The component $\Omega-\Omega_{\rm p}$ is the orbital
frequency of the star in the frame rotating with pattern speed
$\Omega_{\rm p}$.  In this frame, the orbit completes $m$ radial
oscillations for every $l$ azimuthal oscillations.  As $m$ and $l$ are
integers, the resonant orbit closes in the rotating frame.  The
integer $m$ also describes the multiplicity of the pattern the closed
orbit traces out -- the orbit will have $m$-fold symmetry after $l$
rotations.  We note here that, outside of CR, the orbit proceeds in an
opposite sense to the pattern with $\Omega_{\rm p}$.  For this reason,
we set $l<0$ outside CR.  With this terminology, the Outer LR (OLR)
corresponds to the $m:l$ = 2:-1 resonance where $\Omega-\Omega_{\rm
  p}$ is necessarily negative, so by keeping $l<0$ in this region
(beyond CR) Equation \ref{eqn:resCond} is satisfied.  At the OLR then,
an orbit completes 2 radial oscillations for every one azimuthal
oscillation when viewed in the rotating frame with angular frequency
$\Omega_{\rm p}$.

The study of dynamics near resonances has a long history
\citep[e.g.,][]{Contopoulos1970,Lynden-Bell1972,DLB1973}.  The
classical approach involves perturbations to an axisymmetric
model~\citep[see][]{Lichtenberg1983}.  This line of inquiry gives
valuable insights into the dynamics of orbits under small
perturbations and is often used in studying linear stability
\citep[e.g.,][]{Gerhard1991,Palmer1994}.  The addition of a
perturbation series however encounters a problem near resonances.  It
involves the addition of a component with a divisor equal to
$l\kappa-m(\Omega-\Omega_{\rm p})$, which of course approaches zero at
a resonance - the well-known ``problem of small divisors" \citep[see
  e.g.,][]{Arnold1963}.  This results in a divergence of the solution
near a resonance and so does not give a formal answer.

The analysis of resonances in N-body simulations is usually handled by
means of a frequency analysis.  This requires either measuring the
instantaneous frequencies of particles in the simulation or
integrating the chosen orbit in a frozen potential that corresponds to
the density distribution at some timestep in the simulation.  Since
the emergence of large N-body simulations, this method has become
ubiquitous and has proved to be powerful and enlightening.  However,
the method assumes that angular momentum exchange is minimal and has
no significant effect over the course of the orbit. This may lead to
errors if angular momentum exchange is both constant and significant.
In a potential with a significant perturber, instantaneous frequencies
do not give give an accurate representation of the full orbit.  This
is especially true in the case of resonant orbits in discs with large
bars where angular momentum is constantly, and periodically, lost and
gained.  For this reason, the azimuthal frequency must be calculated
over the whole orbit, preferably over one rotation in the rotating
frame.  Below we outline a simple method to get a robust estimate of
$\Omega$ (Section \ref{Sec:Freqs}) using \textit{only} the simulation
output.

A Fourier analysis is often employed to extract frequencies and is
usually sufficient, but in some cases encounters a problem in the
outskirts of the disc where frequencies are low.  A slowly changing
potential can introduce low frequency radial modes and so a lower
cutoff is introduced to remove these, but may also remove some orbits
in the very outskirts of the disc \citep{Ceverino2007}.  As we shall
see below (Section \ref{Sec:ResClump}), the potentials of N-body
simulations rarely have the characteristic of being time independent.
This means that integrating an orbit in a frozen potential will fail
to replicate the real N-body trajectory.  Also, it is well known that
angular momentum exchange between a galactic bar and the halo or a
gaseous component to the disc can vary the pattern speed and strength
of the bar \citep{Athanassoula2002,Athanassoula2003}.  By choosing a
time-independent pattern speed, orbits integrated in frozen potentials
loose this characteristic of realistic barred potentials.

Below we introduce a new method for analysing N-body simulations.  Its
main aim is to uncover resonant orbits in stellar discs using the fact
that, in a rotating frame, the orbit should close and return to a
previously occupied region of phase space.  We outline the method in
Section \ref{sec:PDM}.  The method requires little \textit{a priori}
information.  It only requires the re-calculation of the N-body orbits
in many different frames.  This approach is then applied to the
outskirts of the N-body simulation described by \citet{Shen2010} in
Section \ref{Sec:Applic}.  As a diagnostic tool, we will construct
Fourier spectrograms for the disc (Section \ref{Sec:Fourier}) and
estimate the radial and azimuthal frequencies (Section
\ref{Sec:Freqs}).  This is not required to extract a sample of
resonant orbits from the simulation, but only aids in their
interpretation.  We then explore some of the properties of our
extracted sample of resonant orbits in Section \ref{Sec:Props} and
discuss our findings in Section \ref{sec:Discuss}.

\section{Phase-space Distance Metric}\label{sec:PDM}

\begin{figure}[t]
\begin{center}
\includegraphics[width=0.5\textwidth]{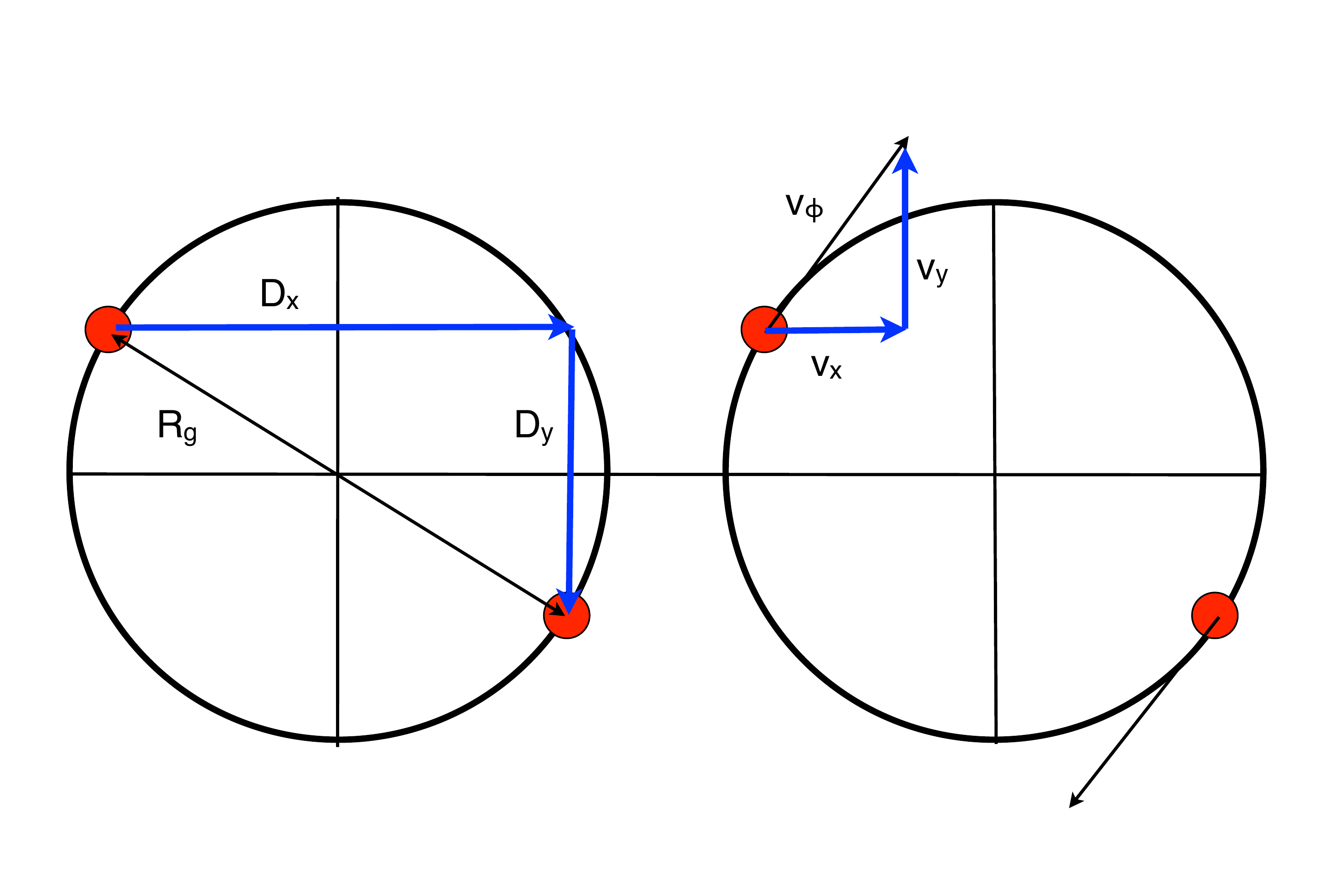}
\caption{Circular orbits suggests a normalisation for our phase space
  distance metric.  The maximum separation on the $(x,y)$-plane
  between two circular orbits differing only in orbital phase is twice
  the guiding radius (left). Using this as our normalisation means the
  spatial component of the phase space distance varies between 0 and
  1.  In a similar fashion, the maximum separation of the kinematic
  component is twice the rotational velocity (right).  This
  normalisation forces the kinematic component of the phase space to
  vary between 0 and 1 also.}\label{fig:Norm}
\end{center}
\end{figure}

In a particular rotating frame, a resonant or periodic orbit forms
a closed pattern and returns to a previously occupied patch of phase
space.  Using this, we make a blind search for orbits that return to
some arbitrarily selected starting position in phase space.  To do
this, we re-calculate the orbits from our N-body simulation in many
different rotating frames.  We define a phase space distance metric,
which measures the distance traveled in the rotating frame $D_{\rm
  ps}$.  A resonant orbit closes in the rotating frame of its
perturber and so $D_{\rm ps}$ for this orbit approaches zero.  In
practice, $D_{\rm ps}$ will not return exactly to zero so we define
some cutoff point, below which the orbit has closed, or nearly so.

\begin{figure*}[t]
\begin{center}
\leavevmode
\includegraphics[angle=90,width=\textwidth]{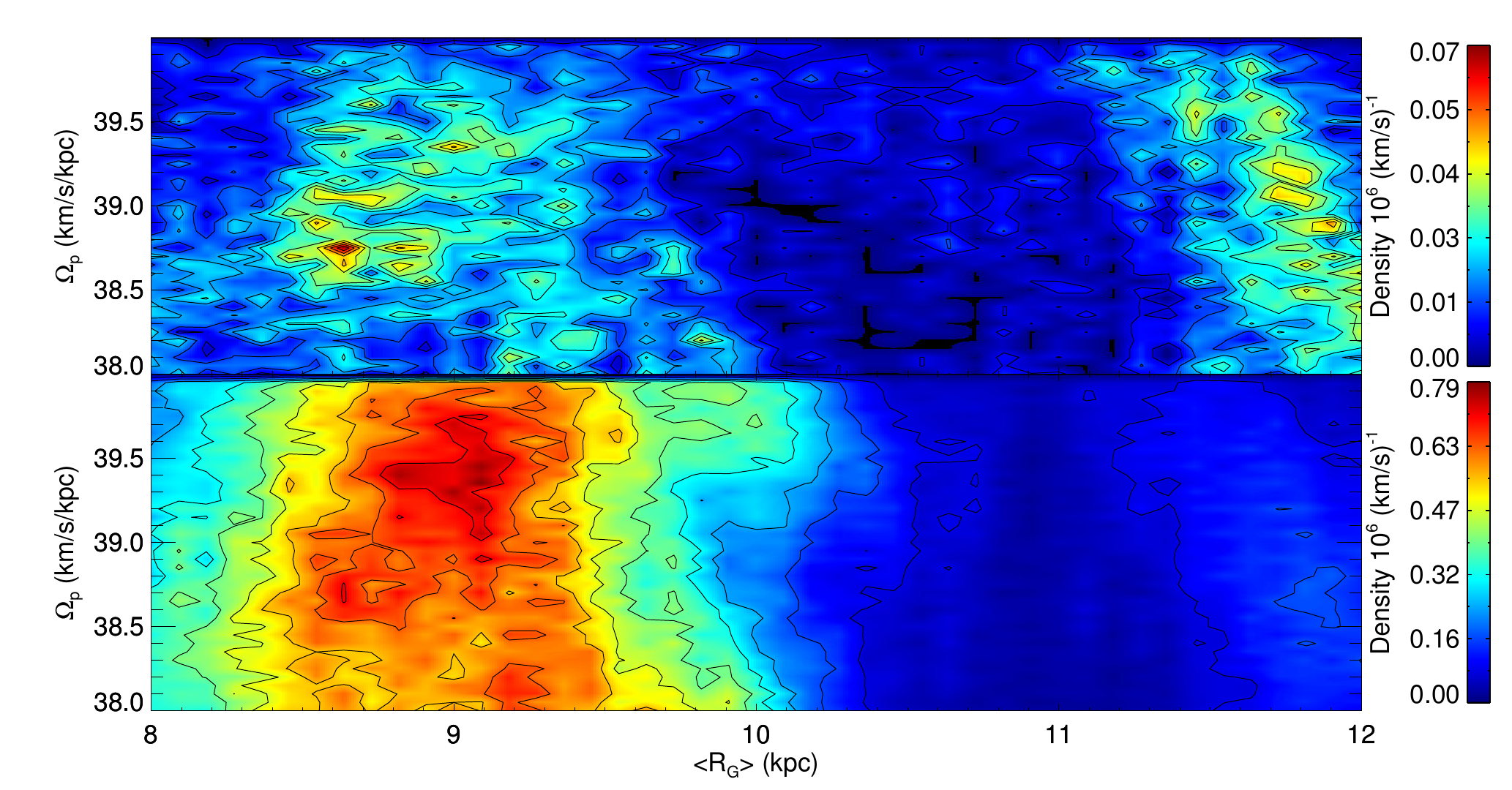}
\caption{Density of closed orbits as a function of their average
  guiding radius and the pattern speed for which they close.  We study
  various rotating frames with frequencies in the range
  $38<\Omega_{\rm p}<40$ km s$^{-1}$ kpc$^{-1}$ using a grid of
  $\triangle\Omega_{\rm p}=0.1$ km s$^{-1}$ kpc$^{-1}$.  The top panel
  indicates our selection when the minimum phase space achieved is
  less than $D_{\rm ps}<0.025$, and the bottom when $D_{\rm ps}<0.04$.
  Most of the closed orbits have $8<R_{\rm g}<10$ kpc with a smaller
  contribution from orbits with $R_{\rm g}>11
  kpc.$}\label{fig:ClosedDens}
\end{center}
\end{figure*}

N-body simulations give the full orbital history in the inertial frame
($R,\phi,z,v_{R},v_{\phi},v_{z}$), which can be recalculated in a
rotating frame ($R',\phi',z',v'_{R},v'_{\phi},v'_{z}$).  We first
define our (arbitrarily chosen) starting position $t_{0}$ and from
there calculate the positions and velocities of particles along their
orbits in many rotating frames using a grid of pattern speeds.  Since
the radial position and velocity are invariant in frames rotating
about the $z$-axis, the calculation of the orbit in the rotating frame
is trivial. We can write

\begin{equation}
\label{ }
\begin{aligned}
R'_{i}&=R_{i} \\
v'_{R,i}&=v_{R,i} \\
\phi'_{i}&=\phi_{i}-i\Omega_{\rm p}\triangle t \\
v'_{\phi,i}&=v_{\phi,i}-R_{i}\Omega_{\rm p}
\end{aligned}
\end{equation}
where $\Omega_{\rm p}$ is the frequency of the rotating frame,
$\triangle t$ is the duration of the timestep (from the N-body
simulation, $\triangle t=0.96$ Myr) and $i$ is the number of timesteps
away from the starting point of the calculation.  To make a
meaningful normalisation of phase space distances, we convert the
positions and velocities from polar to Cartesian coordinates
($[R',\phi',v'_{R},v'_{\phi}]\to[x',y',v'_{x},v'_{y}]$).  For each
particle, the distance along each axis of phase space at timestep
$t_{i}$ is given by
\begin{equation}
\label{ }
\begin{aligned}
D_{x,i}&=|x'(t_{0})-x'(t_{i})| \\
D_{y,i}&=|y'(t_{0})-y'(t_{i})| \\
D_{v_{x},i}&=|v'_{x}(t_{0})-v'_{x}(t_{i})| \\
D_{v_{y},i}&=|v'_{y}(t_{0})-v'_{y}(t_{i})|. 
\end{aligned}
\end{equation}
Formally, phase space is not a metric space and so any distance
measure is arbitrary and tailored to suit the problem under study.
Our general rule of thumb is to first define the expected limits of
the orbit in each dimension and normalise accordingly.  In this case,
we use expectations from simple circular disc orbits.  We note that
for almost circular orbits
\begin{equation}
\label{ }
0\lesssim\sqrt{D^{2}_{x}+D^{2}_{y}}\lesssim2R_{\rm g}
\end{equation}
where $R_{\rm g}$ is the guiding radius of the orbit (i.e. the radius of a
circular orbit with an equivalent angular momentum). This suggests the
following normalisation on the spatial component of the phase space,
\begin{equation}
\label{ }
D_{{\rm p},i}=\dfrac{\sqrt{D^{2}_{x,i}+D^{2}_{y,i}}}{2\bar{R}_{\rm g}}.
\end{equation}
Since angular momentum exchange is ongoing in the disc, we use the
average guiding radius over the course of the orbit, $\bar{R}_{\rm g}$
(note that the guiding radius is evaluated by equating the angular
momentum, $L_{z}$, of the particle with the angular momentum of a
rotation curve derived from azimuthally-averaged force calculations).
However, the median guiding radius may serve as a better normalisation
if sampling is far from uniform along the orbit.  If an orbit near
the centre of the disc is sampled with a constant $\triangle t$, then
there will be significantly more points around apocentre than
pericentre, skewing the value of $\bar{R}_{\rm g}$ to higher values.
By using either the average or median $R_{\rm g}$, the spatial
component of $D_{\rm ps}$ varies between roughly 0 and 1 (see Figure
\ref{fig:Norm}).  We use a similar method to normalise the kinematic
component of the phase space distance.  For almost circular orbits, it
is true that
\begin{equation}
\label{ }
0\lesssim\sqrt{D^{2}_{v_{x}}+D^{2}_{v_{y}}}\lesssim2v'_{\phi}
\end{equation}
so we can calculate a locally normalised distance using
\begin{equation}
\label{ }
D_{v,i}=\dfrac{\sqrt{D^{2}_{v_{x},i}+D^{2}_{v_{y},i}}}{2\bar{v}'_{\phi}}.
\end{equation}
Here, we have again taken an average of $v_{\phi}$ in the rotating
frame over the course of the orbit (a median value may be more
appropriate in the inner parts of the disc).  As with the spatial
component, using $\bar{v}'_{\phi}$ means that the kinematic component
of $D_{\rm ps}$ varies between roughly 0 and 1 (see Figure
\ref{fig:Norm}).  The full phase space distance is then calculated as
\begin{equation}
\label{ }
D_{{\rm ps},i}=\sqrt{D^{2}_{{\rm p},i}+D^{2}_{{\rm v},i}}
\end{equation}
where $D_{\rm ps}$ varies roughly between 0 and $\sqrt{2}$. 

\begin{table}[t]
  \centering 
  \begin{tabular}{c c c c}
\hline \hline
  Cut ($D_{\rm ps}<$ Cut) & Total & $R\le5$ kpc & $R>5$ kpc \\
 \hline
 No Cut & 977,357 & 680,005 & 297,352 \\
  0.1 & 529,015 & 424,097 & 104,918 \\
  0.08 & 425,788 & 355,854 & 69,934 \\
  0.06 & 297,186 & 255,247 & 41,939 \\
  0.04 & 141,061 & 121,747 & 19,314 \\
  0.025  & 40,744 & 35,478 & 5,266 \\
\hline
\end{tabular}
\caption{By imposing stricter cuts on the minimum phase space distance
  $D_{\rm ps}$, we extract samples of cleaner, more strictly closed,
  orbits.}\label{tab:numbers}
\end{table}

This normalisation does introduce separate biases on the spatial and
kinematic components.  On the spatial plane, the normalisation is
biased in favour of particles that gain angular momentum over the
course of the orbit.  A significant gain in angular momentum gives
rise to an increase in the guiding radius, thereby lowering the
spatial phase space distance ($D_{p}$) measured.  In a similar vein,
an increase in ${v}'_{\phi}$ gives rise to a lower kinematic phase
space distance.  We have checked that this bias is weak for reasonable
values of $D_{\rm p}$ and only becomes important when $D_{\rm p}$ is a
significant fraction of the guiding radius ($D_{\rm p}\ge R_{\rm
  g}/2$) -- at this point the orbit will have been excluded by any cut
made on $D_{\rm ps}$.  In any case, resonant orbits experience
periodic changes in angular momentum as they complete their orbits, so
for closed orbits, the variations cancel.

By measuring the lowest phase-space distance of each particle, we can
garner a sample of closed orbits by making a cut below which the
orbits are deemed to have closed.  In order to diminish the effects of
using an arbitrary starting point, which can give spurious results, we
use three different starting points ($t_{0}$).  We only accept orbits
that have a minimum phase-space distance below our predefined cut
using all three starting points.  Each starting point is separated by
200 timesteps or $\approx200$ Myr, which constitutes a significant
fraction of the orbital period.  Using this separation means we choose
very different orbital phases as our starting points, except of course
for those with $T'_{\phi}\approx$ 200 Myr.

\section{An application to an N-body simulation}\label{Sec:Applic}

We test this method using the N-body simulation described by
\citet{Shen2010}.  This simulation begins as an axisymmetric disc
which, due to dynamical instabilities
\citep[see][]{Hohl1971,Kalnajs1978,Toomre1981,Sellwood1981}, develops
a strong bar that persists for $\sim$4.5 Gyr.  The pattern speed of
the bar remains relatively stable at $\sim$38.5 km s$^{-1}$ kpc$^{-1}$
for the duration of the simulation, since there is no gaseous disc
component or live halo to which angular momentum can be deposited.
The amplitude of the bar in this case is expected to result in a
strong resonant reaction in the disc.  The bar also drives recurring
transient spirals.  This persistent strong bar gives us the
opportunity to test the above method in a situation where angular
momentum exchange is significant and long lived, i.e., where a
perturbation series and a frequency analysis may have difficulty.  The
corotation radius (CR) of the bar is at $\approx4.5$ kpc and the outer
Lindblad resonance (OLR) lies in the range $8<R<9$ kpc.

The starting configuration of the simulation is an axisymmetric disc
embedded in a logarithmic halo with a disc scalelength of $h_{R}=1.9$
kpc and scaleheight of $h_{z}=0.2$ kpc.  The disc is comprised of
$10^{6}$ stellar particles with a total mass of $M_{\rm
  D}=4.25\times10^{10}$ $M_{\odot}$, while the mass of the halo is
determined by an asymptotic rotational velocity of $v_{M}=$ 250 km
s$^{-1}$.  The simulation was tailored to match the kinematics of the
Galactic bulge as observed by BRAVA \citep[Bulge RAdial Velocity
  Assay;][]{Rich2007,Howard2008} and this was used to constrain the
angle of the bar with respect to the Galactic centre - Solar position
line.  The final configuration of the system consisted of a bar with a
half-length of 4 kpc and an axial ratio of $\approx0.5$, which is
consistent with previous estimates from gas dynamics
\citep{Englmaier1999,Weiner1999}.  The final bar angle was
$\sim$20\textdegree\ again in agreement with previous studies
\cite[e.g.,][]{Stanek1997,Bissantz2002}.  It has also been recently
shown that the simulation is in good agreement with observations of
Red Clump (RC) stars towards the Galactic centre.  An X-shaped
structure that results from the buckling of the bar \citep[see
  e.g.][]{Raha1991} leads to a bimodal distribution of distances (or
magnitudes) for some fields towards the Galactic centre
\citep{Saito2011} and is well replicated by both this simulation
\citep{Li2012} and that of \citet{Gardner2014}.  In a forthcoming
paper \citep{Molloy2014b}, we isolate the resonant orbits that
contribute to this feature, describing their kinematic properties.

\begin{figure*}[t]
\begin{center}
\leavevmode
\includegraphics[width=\textwidth]{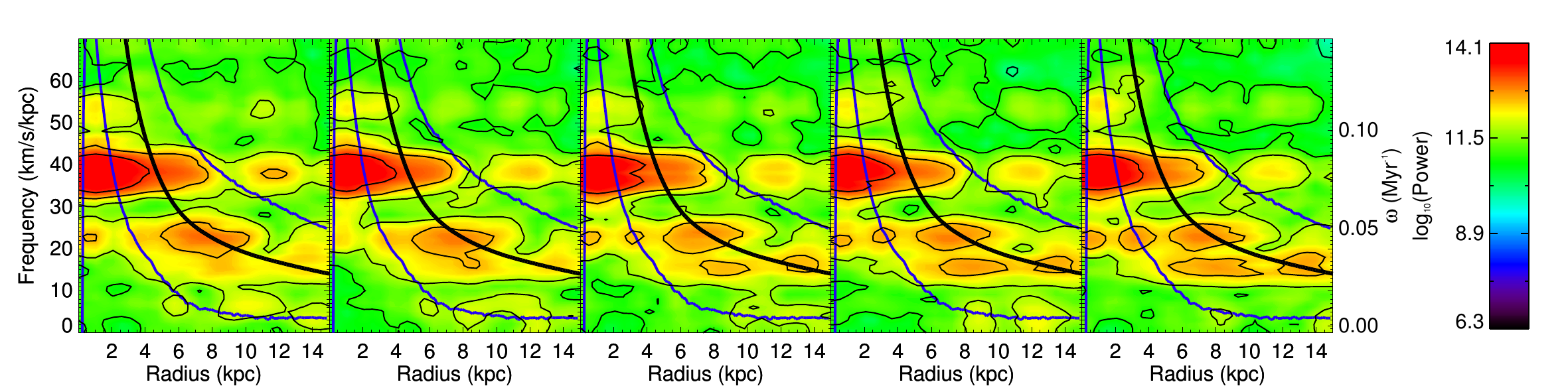}
\caption{$m=2$ Fourier spectrograms spanning $T_{2}-T_{1}=1$
  Gyr.  From left to right, the spectrograms probe the time range from
  2 $\rightarrow$ 3 Gyr, 2.5 $\rightarrow$ 3.5 Gyr, 3 $\rightarrow$ 4
  Gyr, 3.5 $\rightarrow$ 4.5 Gyr and 4 $\rightarrow$ 5 Gyr.  In
  constructing our spectra, we sample the disc every $\sim$10 Myr and
  use a Hanning window function to bracket $T_{1}$ to
  $T_{2}$ (Equation \ref{eqn:PowerSpec}).  The colourbar denotes a log
  scale in arbitrary units.  We also indicate CR with the black
  line and the 2:$\pm$1 LR with the blue lines (Equations \ref{eqn:kapc} and \ref{eqn:LRc}).  
  The most prominent pattern is that of the bar which is rotating with a pattern speed of
  $\sim$38.5 km s$^{-1}$ kpc$^{-1}$.  It extends from the centre of
  the disc out to its CR at $\sim$4.5 kpc.  A two-armed spiral
  emanates from the ends of the bar which also overlaps with a slower
  spiral feature at $\sim$25 km s$^{-1}$ kpc$^{-1}$.  We also see the
  emergence of a transient feature at $\sim$15 km s$^{-1}$ kpc$^{-1}$.}\label{fig:PSpec}
\end{center}
\end{figure*}

We apply the method over many epochs during the lifetime of the
simulation, finding a good sample of closed orbits at all times after
$\sim$2 Gyr.  We focus here on the last Gyr of the simulation
($\sim$4-5 Gyr).  As we are particularly interested in orbits resonant
with the bar, we limit out search to pattern speeds close to that of
the bar ($38\le\Omega_{\rm p}\le40$ km s$^{-1}$ kpc$^{-1}$).  In
Figure \ref{fig:ClosedDens}, we plot the density of closed orbits as a
function of their average guiding radii and the pattern speed of the
rotating frames in which they close.  In the upper panel, we impose a
cut of $D_{\rm ps}<0.025$ and in the lower, a cut of $D_{\rm
  ps}<0.04$. By imposing a tighter cut on $D_{\rm ps}$, we extract a
``cleaner" sample of resonant orbits. As the cut approaches zero, we
converge on the ``parent" orbit for each resonance.  Looser cuts then
encompass orbits that librate about the parent orbit.  In Table
\ref{tab:numbers}, we list the number of resonant orbits extracted for
each cut.  There are many more resonant orbits within CR.  The choice
of cut therefore depends on the region in question.  

Here, we extract a sample of closed orbits whose distribution of
guiding radii peaks at $\sim$9.0 kpc and close in a rotating frame
close to the pattern speed of the bar ($\sim$38.5 km s$^{-1}$
kpc$^{-1}$).  In order to measure the pattern speed, we compute the
numerical derivative of the phase angle of the $m=2$ Fourier component
within CR over a number of timesteps.  The pattern speed of the bar
remains constant at $\approx$ 38.5 km s$^{-1}$ kpc$^{-1}$ for most of
the duration of the simulation (see also Section \ref{Sec:Fourier}).

A source of ambiguity remains, however.  An orbit can close in many
frames. Specifically, an orbit with epicyclic ($\kappa$) and orbital
frequency ($\Omega$) closes in a frame $\Omega_{m:l}$ for which
\begin{equation}
\label{eqn:omegml}
\Omega_{m:l}=\Omega-\left(\dfrac{l}{m}\right)\kappa
\end{equation}
Here, the subscript $m:l$ indicates that the orbit closes with $m$
radial oscillations for every $l$ azimuthal oscillations in that
frame.  When we reproduce Figure \ref{fig:ClosedDens} over a much
wider range of pattern speeds ($25\le\Omega_{\rm p}\le55$ km s$^{-1}$
kpc$^{-1}$), we see multiple overdensities corresponding to this
degeneracy.  So, while we have limited our search for resonant orbits
associated with the bar, we may also be extracting orbits that are
driven by slower or faster moving patterns and which also happen to
close in a frame close to that of the bar's.  To clarify the matter,
we measure the orbital and epicyclic frequencies for our sample of
closed orbits.  Using this information, we determine the pattern
speeds of the rotating frames in which our orbits will close for
various families of orbits (e.g., $m:l=$ 1:-1, 1:-2, etc.).

\subsection{Measuring Frequencies}\label{Sec:Freqs}

It is difficult to estimate reliably the orbital frequencies of stars
in discs in which angular momentum exchange is persistent and
significant.  Taking instantaneous measurements of the azimuthal
frequency $\Omega$ introduces an error since the angular momentum at
one point in the orbit may be, sometimes significantly, different from
that at another point.  For this reason, it is better to estimate the
frequency over the course of a full orbit.  This is especially true
for resonant orbits -- the periodic loss and gain of angular momentum
is averaged out over some integer number of azimuthal oscillations in
the rotating frame.

Our method makes use of the fact that an orbit plotted in a rotating
frame with the same frequency as the orbital frequency of its guiding
centre will trace out an ellipse (its epicycle) which appears
stationary.  In this frame, the epicycle doesn't precess around the
centre of the disc and explores only a limited range in $\phi$.  In
each rotating frame, the orbit has a minimum and maximum azimuth
($\phi_{\rm min}$ and $\phi_{\rm max}$).  The range in $\phi$
traversed by the orbit is $\bigtriangleup\phi=|\phi_{\rm
  max}-\phi_{\rm min}|$.  The method minimises $\bigtriangleup\phi$ by
scanning various rotating frames with higher and higher resolution
until the variation between adjacent frames reaches some predefined
limit.  For well-behaved orbits, this converges to the true orbital
frequency of the guiding radius.  The epicyclic frequency is extracted
from the period of the radial oscillation.

\subsection{Fourier Components}\label{Sec:Fourier}

Discerning which of the non-axisymmetric features of the disc (e.g.,
the central bar or spiral arms) resonate with our closed orbits
requires an analysis of the wave patterns that propagate through
the disc.  For this, we decompose the disc into its Fourier components.
We follow the procedure outlined by \cite{Quillen2011} to construct
our spectrograms.  For equally spaced radial bins at each timestep, we
measure
\begin{equation}
\label{ }
\begin{aligned}
W_{\rm C}(r,t)&=\displaystyle\sum_{j}\cos(\phi_{j}) \\
W_{\rm S}(r,t)&=\displaystyle\sum_{j}\sin(\phi_{j})
\end{aligned}
\end{equation}
where we sum over all $j$ particles in each radial bin. We then
numerically integrate over the complex function
\begin{equation}
\label{eqn:PowerSpec}
\tilde{W}(\omega,t)=\int^{T_{2}}_{T_{1}}\exp(-i\omega t)h(t)[W_{\rm C}(r,t)+iW_{\rm S}(r,t)]dt
\end{equation}
where we sample over the frequency domain 0 $<\omega<$ 70 km s$^{-1}$
kpc$^{-1}$.  We also use a Hanning window function $h(t)$ that spans
the time window $T_{1}$ to $T_{2}$.  We construct spectrograms using
the $m$ = 1, 2, 3 and 4 Fourier components for five different time
windows with $ T_{2}-T_{1}=0.96$ Gyr (1,000 timesteps) from $\sim$2 to
$\sim$5 Gyr (i.e. 2 Gyr $\rightarrow$ 3 Gyr, 2.5 Gyr $\rightarrow$ 3.5
Gyr, 3 Gyr $\rightarrow$ 4 Gyr, etc.).  The spectrograms are sampled
every 9.6 Myr (10 timesteps) so that we can uncover transient features
(such as spiral arms) that require a high temporal resolution.

The $m$ = 1 $\&$ 3 spectrograms exhibit patterns many orders of
magnitude weaker than the $m$ = 2 patterns, whilst the $m$ = 4
spectrograms show only the first harmonic of the $m$ = 2 patterns.  We
give only the $m$ = 2 spectrograms in Figure \ref{fig:PSpec} with the
vertical axis indicating the pattern speed and the horizontal axis the
radial extent of the features.  The colour shows the strength of the
feature (in arbitrary units and on a log scale).  We show CR as the
black line and also the 2:$\pm$1 LR as blue lines.  The $m:l$ LR are
derived from the rotation curve calculated using force measurements
around the disc.  This gives $\Omega_{\rm c}(R)$, the azimuthal
frequency of circular orbits at $R$.  Using a numerical
differentiation of $\Omega_{\rm c}^{2}$, we derive $\kappa_{\rm c}$
from
\begin{equation}
\label{eqn:kapc}
\kappa^{2}_{\rm c}(R)=R\dfrac{d\Omega_{\rm c}^{2}}{dR}+4\Omega_{\rm c}^{2}.
\end{equation}
The $m:l$ Lindblad resonance (LR) is then (where, as before, $l<0$ for
$R$ exceeeding the corotation radius)
\begin{equation}
\label{eqn:LRc}
LR_{m:l}(R)=\Omega_{\rm c}-\dfrac{l}{m}\kappa_{\rm c}.
\end{equation}

\begin{figure*}[!t]
\begin{center}
\leavevmode
\includegraphics[angle=90,width=\textwidth]{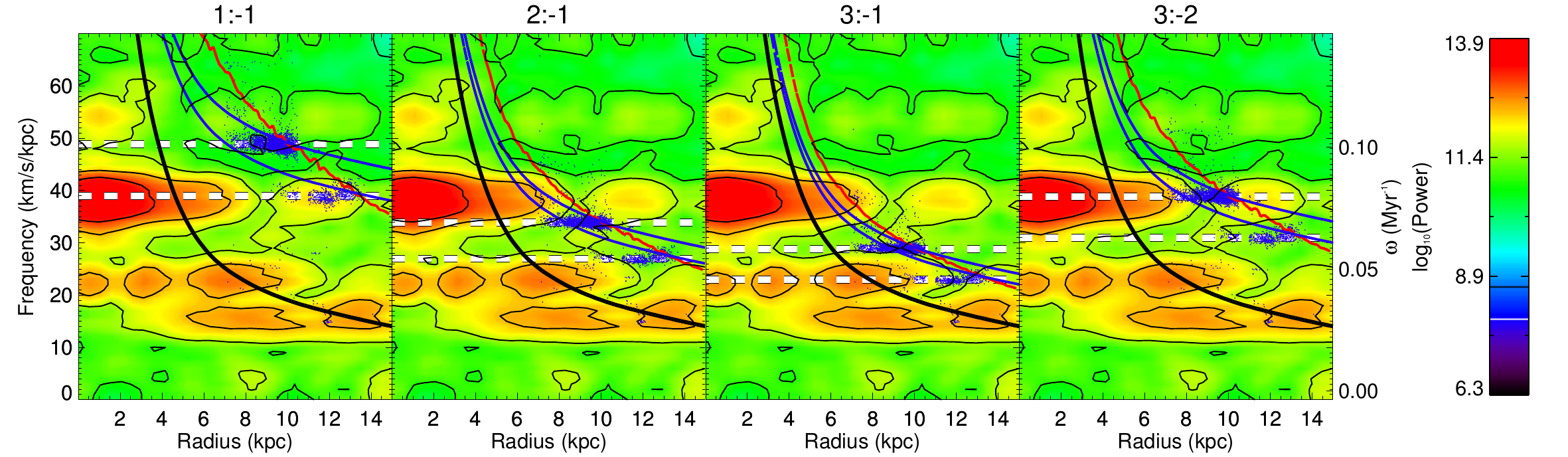}
\caption{The $m=$ 2 component spectrogram covering the time window 4
  $\rightarrow$ 5 Gyr, i.e. the right-most plot in Figure
  \ref{fig:PSpec}.  Each plot shows the \textit{same} spectrogram with
  the CR indicated as the black line.  We indicate the $m:l$ LR as the
  blue and red lines -- the 1:-1 LR in the first plot, the 2:-1 LR in
  the second plot, and so on.  The red line uses the epicyclic
  approximation (i.e. $\kappa$ is derived from the rotation curve,
  $\Omega_{\rm c}$), while the blue lines use the median $\kappa$ from
  samples of orbits.  Specifically, the outer blue line uses the
  median $\kappa$ from the particles in the range $6<R<11$ kpc, while
  the inner blue line uses the median $\kappa$ from particles with
  $R>11$ kpc.  In each plot, we overlay the distributions of
  $\Omega_{m:l}$ (Equation \ref{eqn:omegml}) for particles with
  $D_{\rm ps}<0.25$ as the blue points -- they represent the
  \textit{same} particles, the only difference being the frequency of
  the rotating frames in which they close as 1:-1 (first), 2:-1
  (second), 3:-1 (third) $\&$ 3:-2 (fourth) orbits.  The median
  frequency of each $\Omega_{m:l}$ distribution is plotted as the
  horizontal dashed line.  Two separate groups appear at distinct
  frequencies.  The larger group (with higher frequencies) is
  coincident with the 3:-2 LR and is in good agreement with the
  pattern speed of the bar as a family of 3:-2 orbits.  The lower
  frequency group could be resonating in a 1:-1 fashion with the
  strong bar pattern or in a 3:-1 fashion with a weaker pattern at
  $\sim$22 km s$^{-1}$ kpc$^{-1}$.}\label{fig:PSpecOverplot}
\end{center}
\end{figure*}

The spectrograms of the $m=2$ Fourier components indicate the radial
extent and the rotational frequency of two-armed structures, like a
bar or spiral arms, in the disc.  The most prominent feature is the
bar pattern, which extends from the centre of the disc out to its
corotation radius, rotating at $\sim$40 km s$^{-1}$ kpc$^{-1}$.  A
weaker feature extends from the end of the bar in all $m=2$
spectrograms and reaches as far as the 2:-1 LR.  The orbits that make
up the structure of the bar are prohibited from extending beyond
corotation \citep[see][]{Chirikov1979,Contopoulos1980}, so this
feature is likely due to spirals emanating from the ends of the bar.
We see another two-armed structure that also extends from the end of
the bar, but rotates with a lower frequency ($\sim$22 km s$^{-1}$
kpc$^{-1}$).  At later times, a further spiral is seen at a still
lower frequency ($\sim$15 km s$^{-1}$ kpc$^{-1}$).  This spiral has
its ILR at the same radius as the bar's CR which may indicate the
presence of mode-coupling \citep{Tagger1987,Minchev2012}.  The bar is
extremely stable in this simulation and only slows down very slightly
over the last 4 Gyr.  The spectrograms indicate clearly the presence
of multiple patterns in the disc, which contribute to the ambiguity
about the forcing resonant perturber.

\subsection{Diagnostic}

We now calculate the pattern speeds in which orbits close for various
values of $m$ and $l$ ($\Omega_{m:l}$, Equation \ref{eqn:omegml}).
Using this along with our Fourier spectrograms, we can decipher the
most likely resonant origin for our sample of closed orbits.  We make
two assumptions in the following analysis:
\begin{enumerate}
\item Resonant orbits occur due to non-axisymmetric patterns \textit{inside} the guiding radii of the orbits. 
\item Resonant orbits occur roughly at the location of a Lindblad Resonance/Radius (LR) \citep[see][section 3.3.3]{BT2}.
\end{enumerate}
By plotting our $\Omega_{m:l}$ distributions over our $m=$ 2
spectrograms, we look for the coincidence of the distributions with
non-axisymmetric features in the disc, satisfying our assumptions
above.  We neglect the $m=$ 1, 3 \& 4 spectrograms since the amplitude
of those patterns are dwarfed by the $m=$ 2 components.

Figure \ref{fig:PSpecOverplot} shows the $m=$ 2 spectrograms for the
final Gyr of the simulation (the right-most plot in Figure
\ref{fig:PSpec}).  The spectrograms are identical in each plot.  We
have overlain the distributions of $\Omega_{m:l}$ for the 1:-1, 2:-1,
3:-1 \& 3:-2 (first, second, third \& fourth plots respectively) orbit
families as the blue dots.  The blue dots in each of the plots mark
the \textit{same} particles. We have chosen only the particles that
have $D_{\rm ps}<0.025$, namely the ``cleanest", or ``most-closed",
orbits.  Their different position in each plot represents the
different frames in which they close as different types of orbits.
We identify two separate groups of closed orbits with significantly
different distributions in $\Omega_{m:l}$ -- one lies inside 11 kpc,
while the other lies outside.  For both groups, we have plotted the
median frequency in each distribution (given by the horizontal line).

We also overplot the LR curves (blue/red) along with the CR curve
(black) for the disc (the rotation curve is almost constant over the
duration of the simulation so we use the curves as they are at the end
of the simulation).  The 1:-1, 2:-1, 3:-1 and 3:-2 LR are shown in the
first, second, third $\&$ fourth plots respectively.  For the LR
curves, we have used two different approaches corresponding to the
blue and red lines.  The red LR curves use $\kappa_{\rm c}$ which is
evaluated from the epicyclic approximation using numerical
differentiation.  The blue LR uses the median $\kappa$ from the sample
of resonant orbits along with $\Omega_{\rm c}$.  For the outer (inner)
LR, we use the median $\kappa$ of the orbits inside (outside) 11 kpc
(Note that these curves are only valid over the radius spanned by the
blue points).  Also, the particles generally rotate slower than
indicated by the rotation curve meaning that $\Omega_{\rm c}$ is an
overestimate of the true $\Omega$.  This means that at a given radius
the frequency of the LR is also slightly overestimated.

To begin with, we focus solely on the group with $\overline{R_{\rm
    g}}$ between 6 and 11 kpc.  We look for coincidence of the
blue points (the distributions of $\Omega_{m:l}$) with both a strong
pattern (the red/yellow peaks in the spectrogram) and the location of
a LR (the red/blue lines).  We can see that the distributions of
$\Omega_{m:l}$ (the blue points) only satisfy our assumptions above
for the $m:l=$ 3:-2 orbit family (fourth plot).  In this case, the
guiding radii of the orbits are coincident with the 3:-2 LR and have
almost the same pattern speed as the bar ($\sim$40 km s$^{-1}$
kpc$^{-1}$).  This is good evidence that the group of periodic orbits
inside 11 kpc are resonating in a 3:-2 fashion with the bar.  The
distributions of $\Omega_{m:l}$ are not coincident with any strong
pattern as 1:-1, 2:-1 or 3:-1 orbits.

Similarly, for the group of orbits that lie outside 11 kpc, we see
that they satisfy our assumptions in the $m=2$ spectrogram for the
1:-1 orbit family.  First, as a family of 1:-1 orbits, they close in a
rotating frame with almost the same pattern speed as the bar.
Secondly, their location on the spectrogram coincides with the 1:-1
LR.  There is also some agreement for the 3:-1 orbit family -- a small
feature lies inside the distribution with $\Omega_{\rm p} \approx22$
km s$^{-1}$ kpc$^{-1}$.  However, since the bar is much stronger than
the pattern at $\sim$22 km s$^{-1}$ kpc$^{-1}$, it is likely that
these orbits are resonating in a 1:-1 fashion with the bar.  Only a
handful of particles inhabit this resonance ($\sim$1,750 for $D_{\rm
  ps}<0.04$, $\sim$2,300 for $D_{\rm ps}<0.06$) and they make no
significant contribution to kinematic structures in the disc (even
though they make up $\sim$20\% of all particles with $11<R_{\rm g}<15$
kpc for both cuts), so we do not consider them further.  The 3:-2
orbits make up only $\sim$1\% of all particles with $6<R_{\rm g}<10$
kpc for $D_{\rm ps}<0.025$ but increases to over 30\% for $D_{\rm
  ps}<0.1$.

We remark that the measured frequencies ($\Omega$ and $\kappa$) are
estimates of the true frequencies.  They do however serve well as a
diagnostic and remove any ambiguity regarding the forcing resonant
perturber.  By estimating the frames ($\Omega_{m:l}$) in which our
sample of resonant orbits close as different orbit types, we can
ignore families for which there are no patterns with commensurate
frequencies.  For example, we see that for the 2:-1 orbit family
(second plot), there are no patterns that could force this type of
resonance and so we disregard it.  The best agreement lies with the
bar pattern being the forcing perturber for the 3:-2 and 1:-1 orbit
families.

\section{Resonant Orbit Dynamics}\label{Sec:Props}

\begin{figure}[!t]
\begin{center}
\includegraphics[width=0.45\textwidth]{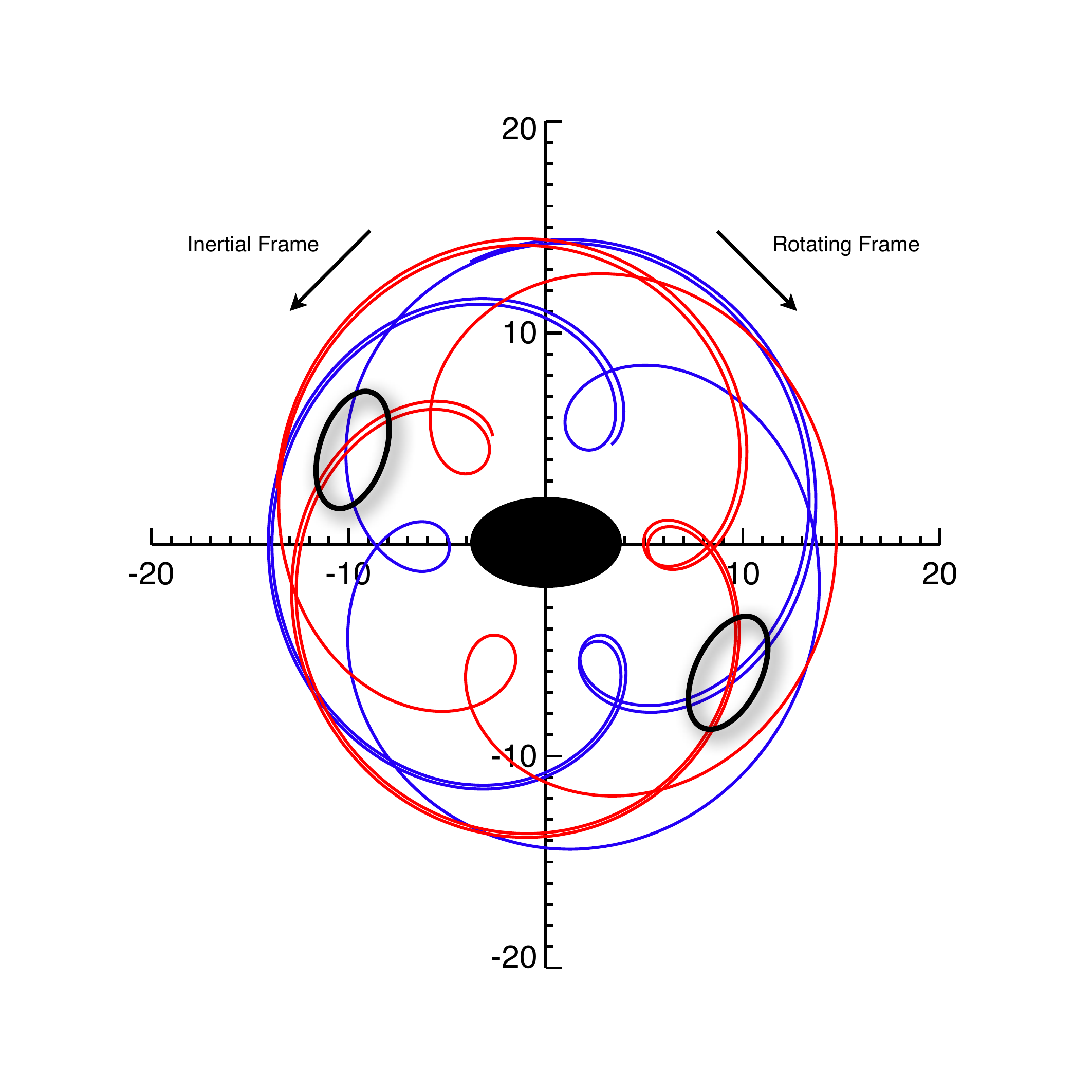}
\caption{Possible orientations of the 3:-2 orbit, which has two-fold
  symmetry with respect to the bar aligned with the $x$-axis.  We
  define a chirality so the the orientation with a pericentre on the
  negative $x$-axis is left-handed (blue) and that with a pericentre
  on the positive $x$-axis is right-handed (red).  The orientations of
  the 3:-2 orbit family allow six possible pericentres (and
  apocentres).  When we overplot the two orientations, we can identify
  locations (with respect to the bar) at which there are both inward
  and outward moving groups (indicated by the black ovals).  In the
  inertial frame, the bar and particles rotate in an anti-clockwise
  fashion. In the rotating frame with $\Omega=\Omega_{bar}$, the bar
  remains aligned with the $x$-axis and the particle move in a
  clockwise fashion.  The locations occur on the trailing side of the
  bar at longitudes similar to the Solar position.}\label{fig:32Orient}
\end{center}
\end{figure}

Our sample of resonant orbits represents kinematic substructure in the
disc that is not present in an axisymmetric potential.  The choice of
a cut on the minimum phase space distance may seem arbitrary and does
introduce uncertainty on the fraction of particles on resonant orbits.
However, it does demonstrate the types of resonances that are
populated and can persist for a significant duration in discs with a
strong central bar.  By extracting a sample of resonant orbits from
N-body simulations, we can also look more closely at their behaviour
as the disc evolves.  Here, we concentrate on resonances beyond CR but
this method works just as well for the bar orbits.

\begin{figure}[t]
\begin{center}
\includegraphics[width=0.5\textwidth]{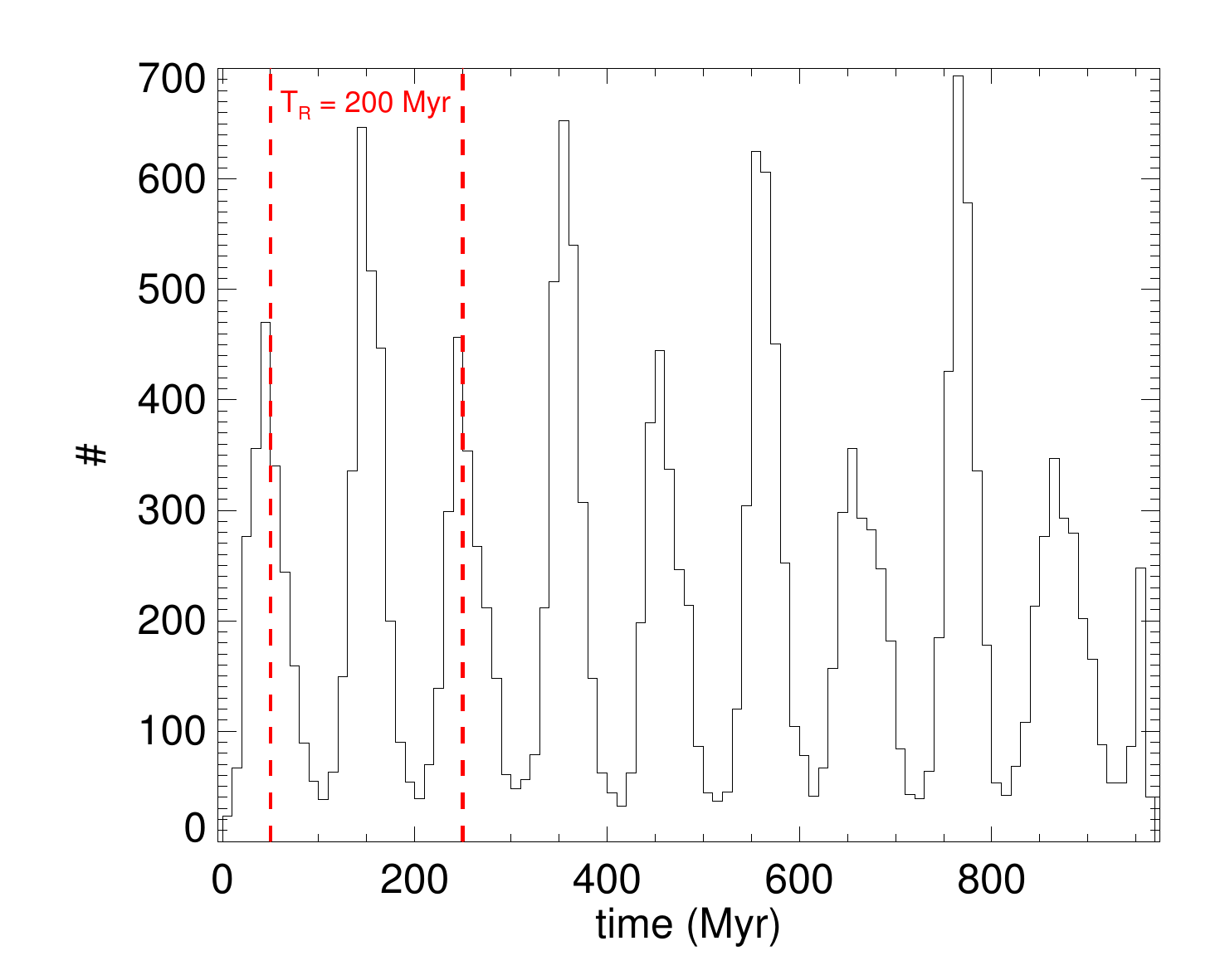}
\caption{The distribution of times at which our closed orbits reach
  their pericentres (over the last 0.96 Gyr). Note that the orbits
  mostly conspire to reach their pericentres at roughly the same time.
  This is the phenomenon of {\it resonant clumping} in the phase angle
  $\theta_{R}$ (where $\dot\theta_{R}=\kappa$).  We indicate the
  epicyclic period, $T_{R}$, for our sample of 3:-2 closed orbits as
  the vertical red lines.  Peaks occur on a timescale of one half of
  $T_{R}$ which means our closed orbits are made up of two
  \textit{groups} whose epicyclic motion is out of phase by
  $\pi$.}\label{fig:PeriTime}
\end{center}
\end{figure}

\begin{sidewaysfigure*}[t]
\begin{center}
\leavevmode
\vspace{8.0cm}
\includegraphics[width=\textwidth]{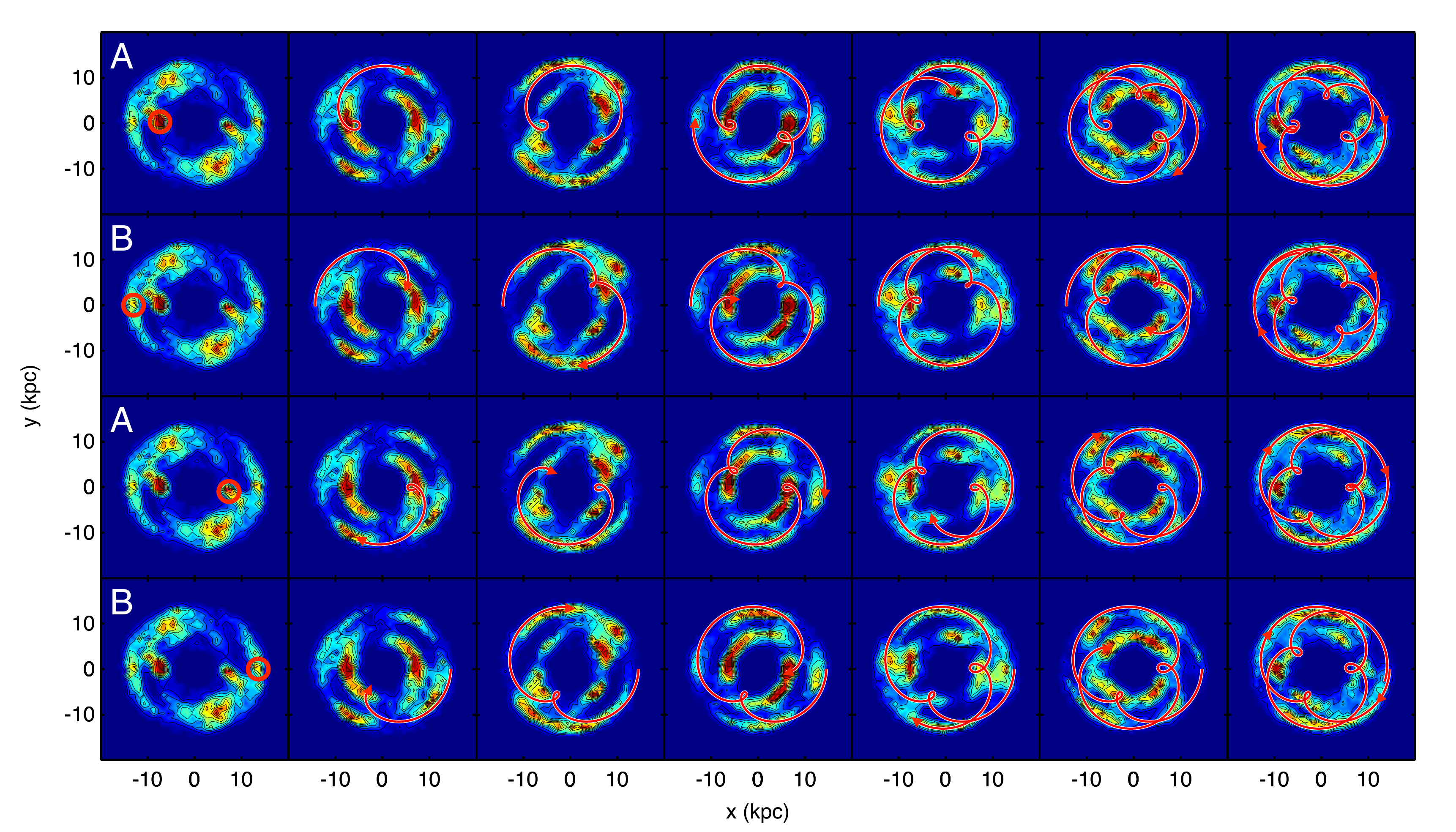}
\caption{Surface density maps of the sample of 3:-2 closed orbits
  only. The long axis of the bar is aligned with the $x$-axis.  In
  each column, the surface density maps are identical.  Successive
  columns show the surface density each time a pericentre or apocentre
  is reached as inferred from the histogram in Figure
  \ref{fig:PeriTime} (every $\sim$100 Myr = $T_{ R}$/2).  In the first
  column, we identify four overdensities that lie on the $x$-axis -
  each row indicates a different one with the red circle.  We define a
  \textit{groups} as those particles whose epicyclic motion is in
  phase, i.e. they reach their pericentres or apocentres at the same
  time.  The overdensities indicated on rows one and three represent
  particles from group A since they both are at pericentre.
  Overdensities from group B are indicated on rows two and four --
  they both reside at apocentre.  We follow the orbital trajectory of
  a particle from each overdensity in the frame rotating with the bar
  -- the orbits all move in a clockwise direction.  Group A orbits
  move from pericentre to apocentre, whilst Group B orbits do the
  converse.  Each of the orbits traces out a 3:-2 orbit pattern with
  the top two rows exhibiting a left handedness (the blue orbit from
  Figure \ref{fig:32Orient}) and the bottom two rows a right
  handedness (the red orbit from Figure \ref{fig:32Orient}). We include
  an animation of the evolution of the 3:-2 orbits as supplementary
  online material.}\label{fig:TPointsxy}
\end{center}
\end{sidewaysfigure*}

\subsection{Resonant Clumping}\label{Sec:ResClump}

Our sample of 3:-2 orbits behave in a highly coordinated fashion.
They are not distributed evenly around the orbit, but instead exist in
multiple populations that vary in their epicycle and azimuthal phase.
In the following, we describe populations that are in phase in
epicycle as \textit{groups}.  Within these groups are populations with
the same azimuthal phase which we call \textit{subgroups}.  With this
nomenclature, the groups all reach their peri-/apocentres at the same
time, while the subgroups have their peri-/apocentres at different
azimuths.  Figure \ref{fig:32Orient} shows the two possible
orientations of the 3:-2 orbit -- each orientation has one pericentre
at either end of the bar which is aligned with the $x$-axis and the
orbits move in a clockwise fashion (in the inertial frame the disc
rotates in an anti-clockwise fashion).  Since we have defined $l$ to
be negative for orbits beyond CR, Equation \ref{eqn:omegml} means that
$\Omega_{m:l}$ is always larger than $\Omega$, the orbital frequency
in the inertial frame (at least for orbits with $\kappa>0$).  The
orbital frequency in the rotating frame, $\Omega'$, is then
necessarily negative, i.e.,
\begin{equation}
\label{ }
\Omega'=\Omega-\Omega_{m:l}<0,
\end{equation}
so that in the rotating frame the orbit moves in an opposite sense to
the inertial frame.  It is useful here to define a chirality, or
\textit{handedness}, for each orientation.  Those orbits that have one
of their pericentres on the negative $x$-axis are left-handed (i.e.,
on the left end of the bar - the blue orbit), while those with one on
the positive $x$-axis are right-handed (the red orbit).

\begin{figure*}[t]
\begin{center}
\includegraphics[angle=90,width=\textwidth]{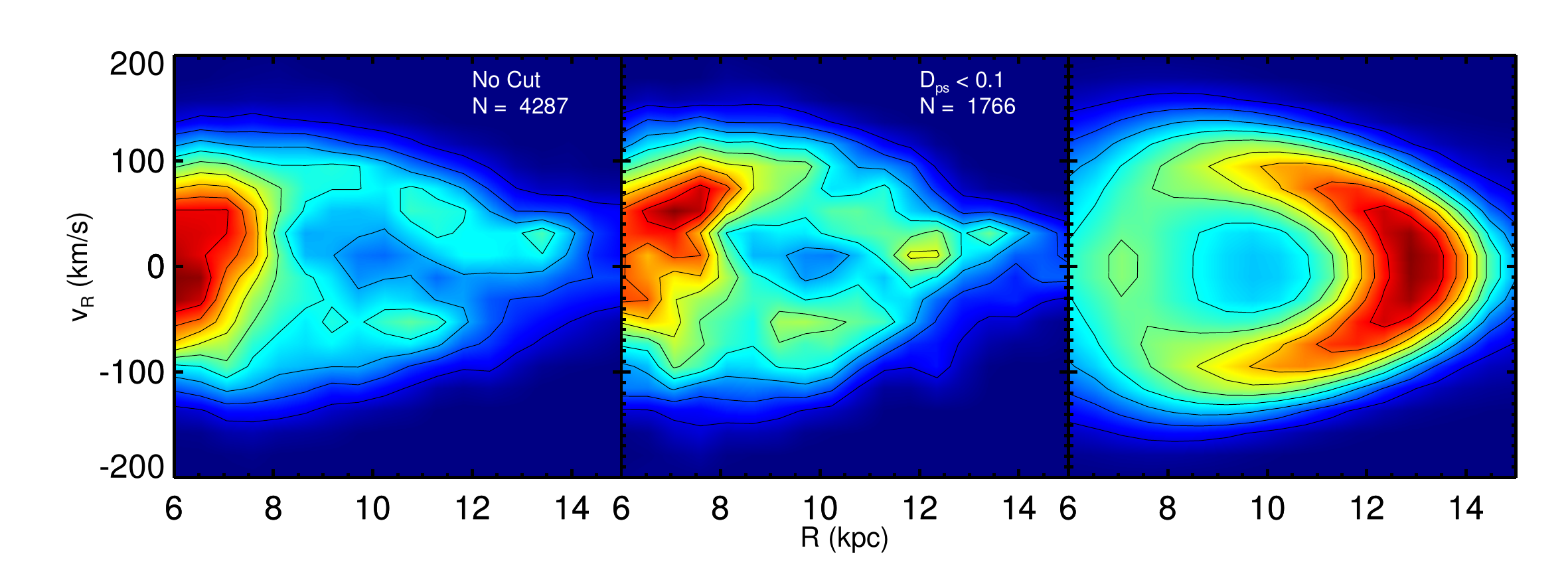}
\caption{The distribution of $v_{R}$ as a function of $R$ for a pencil
  beam towards the anti-Centre at 4.8 Gyr. The number of particles
  $N$ is also indicated.  The disc is rotated so that the bar is at
  an angle of 20\textdegree\ with respect to the $x$-axis.  Particles
  for which $|z|\le0.5$ kpc and $|y|\le1$ kpc make up the
  distributions.  On the left, we show the distribution for
  \textit{all} particles.  A strong kinematic feature is present,
  namely the bimodal distribution of $v_{R}$ between 10 and 12 kpc.
  The middle plot shows the distribution generated by only the
  resonant orbits using a cut on the phase space distance of $D_{\rm
    ps}<0.1$.  The right plot shows the time-averaged imprint of the
  resonant orbits in $R$-$v_{R}$ space.  The density in this plot is a
  measure of the time spent at each location, meaning that the
  persistence of kinematic features scales with the radius at which
  they occur.}\label{fig:vrDenseCut}
\end{center}
\end{figure*}

If the particles populating these types of orbits are distributed
uniformly in azimuth, then we should see bimodal $v_{R}$ distributions
at a number of discrete angles with respect to the bar.  This is not
the case, however.  When we measure the times at which the particles
reach their pericentres, we see that they are arranged in such a way
as to reach their pericentres at almost the same time.  Figure
\ref{fig:PeriTime} shows the distribution of times at which our closed
3:-2 orbits pass through their pericentres.  This is clear evidence
that the particles are not uniformly distributed and so we expect a
\textit{resonant clumping} of the particles in azimuth as all of the
particles coherently reach their pericentres.  It is also clear that
the 3:-2 resonant orbits are split into two distinct \textit{groups}
whose epicycles are out of phase by $\pi$.  The median $\kappa$ for
the 3:-2 orbits is $\approx30$ km s$^{-1}$ kpc$^{-1}$, which equates
to an epicyclic period of $T_{R}\approx205$ Myr.  Figure
\ref{fig:PeriTime} shows that, as one group of 3:-2 orbits reaches
pericentre, the other reaches apocentre.  The groups then pass each
other as they each complete their epicyclic motion -- i.e., they
comprise of an outward moving group and an inward moving group.  This
coordinated movement gives rise to transient overdensities at
pericentre and apocentre and to bimodal distributions in $v_{R}$. 

We have checked that this phenomenon is not some artefact introduced by our use of the phase space distance method. By estimating $\Omega$ and $\kappa$ for a random sample of 10\% of the simulation particles, we extract an independent sample of 3:-2 periodic orbits. This sample also exhibits clumping on timescales consistent with the epicyclic period and confirms that no bias towards certain epicyclic phases is introduced.

Each \textit{group} is host to a number of \textit{subgroups}.  When a
group reaches pericentre, its constituent subgroups populate
over-densities which occur at different azimuths.  The preferred
azimuths have distinct locations with respect to the bar -- the
allowed locations correspond to the six pericentres shown in Figure
\ref{fig:32Orient} (three red and three blue).  It is not necessary
for all six locations to be populated but, in order to avoid
lopsidedness, if one is populated then another on the opposite side
must also be populated with a similar number of particles.  We follow
these coherent motions by plotting the surface density of the 3:-2
resonant orbits at times when a peri/apocentre is reached (Figure
\ref{fig:TPointsxy}).  The surface density maps in each column are
identical and are plotted when a peak occurs in the distribution of
pericentre times in Figure \ref{fig:PeriTime}.  In each map, the bar
is aligned with the $x$-axis.  In the first column, we see four
overdensities that lie along the $x$-axis -- two inner and two outer
marked by the red circles in each row.  The inner overdensities (with
$R\approx$ 8 kpc) correspond to a single group, labelled group A.
Each (inner) overdensity then corresponds to a subgroup of group A.
These subgroups are \textit{in-phase} with respect to their epicycles
but are exactly $\pi$ out of phase in their azimuthal motion.
Similarly, the outer overdensities on the $x$-axis (with $R\approx$ 12
kpc) constitute group B, whose \textit{epicycles} are out of phase by
$\pi$ compared to group A.  As with group A, group B is host to
subgroups -- those subgroups on the $x$-axis are also $\pi$ out of
phase in azimuth.  So, to reiterate, two overdensities in group A are
denoted by the red circles in rows one and three, with both rows
indicating separate subgroups.  Rows two and four indicate two of the
overdensities from group B with, again, each row indicating separate
subgroups.  For each separate subgroup, we choose a particle and
follow its trajectory through successive apo/pericentres.  In all
cases, the bar is aligned with the $x$-axis and the orbits progress in
a clockwise direction.

In the second column, the particles from group A have moved from
pericentre to apocentre, and vice versa for group B.  At some
intermediate point, the particles from each group have passed each
other, with group A having positive $v_{R}$ and group B having
negative $v_{R}$.  This passage generates a bimodal distribution of
$v_{R}$ in the regions of space where the two groups pass each other.
As we move to the third column, group A has returned to pericentre
while group B has moved again to apocentre.  Each successive column
indicates a peri- or apocentre until the final column, where the 3:-2
orbit has been completed.  In terms of chirality, each group is host
to both left- and right-handed 3:-2 orbits (the top two rows are
left-handed and the bottom two right-handed).  Note also that for this
simulation, each group is host to \textit{four} out the six possible
subgroups.  This is evidenced by the absence of an overdensity at the
end of the bar in every third snapshot.  If all six subgroups were
populated, then each time a pericentre is reached, an overdensity
would appear at the ends of the bar.

The evolution of these distinct orbital \textit{groups} and
\textit{subgroups} leads to a ``pulsing" of the density distribution
in the disc with stars periodically and coherently moving inwards and
outwards.  These transient overdensities contribute to the
time-dependent, but periodic, nature of the galactic potential.  The
presence of these two populations contributes to stabilizing what seems
to be an inherently lop-sided configuration.  This mechanism of
coherence among the resonant orbits suggest that strong kinematic
features can be generated as inward and outward moving groups
encounter each other.

\subsection{Kinematic Signatures}

This bimodal distribution of $v_{R}$ constitutes a potentially
observable feature from the Solar position.  At some point,
overdensities appear in line with the long axis of the bar on one side
of the disc, one due to an accumulation at pericentre and one due to
an accumulation at apocentre (one from groups A and B; e.g., rows 1 \&
2, Figure \ref{fig:TPointsxy}).  As they both progress to the extremes
of their respective epicycles, the bar moves ahead in azimuth.  This
means that they pass each other on the trailing side of the bar,
generating a bimodal distribution in $v_{R}$.  This bimodal
distribution appears at distinct azimuths -- specifically, between the
allowed locations for the pericentres and apocentres (see Figure
\ref{fig:32Orient}).  

The angle of the bar with respect to the line joining the Sun and
Galactic Centre is poorly constrained. It is estimated to be in the
region of 10\textdegree\ to
40\textdegree\ \citep[e.g.,][]{Stanek1997,Freudenreich1998,Robin2012,Wang2013,Cao2013}.
For an angle of $\sim$20\textdegree\ on the trailing side of the bar,
there is a bifurcation in the radial velocities between 10 and 11 kpc.
We indicate these regions with the black ovals in Figure
\ref{fig:32Orient}.  We can simulate a field of view towards the
anti-Centre and measure the distribution of $v_{R}$ as a function of
galactocentric distance $R_{\rm GC}$.  By doing this for only the
resonant orbits, we can see how the bimodal nature of the $v_{R}$
distribution emerges (Figure \ref{fig:vrDenseCut}).  If we rotate the
disc so that the bar is at an angle of 20\textdegree\ with respect to
the $x$-axis and select only particles that have $|y|\le1$ kpc and
$|z|\le0.5$ kpc, a bimodal distribution of $v_{R}$ is present between
10 and 11 kpc.  This feature is still significant when we consider all
(i.e. resonant and non-resonant) particles (left, Figure
\ref{fig:vrDenseCut}).

This result is in good qualitative agreement with the recent
observations of \citet{Liu2012a} who, for a sample of Red Clump stars
in a pencil beam towards the Galactic anti-Centre, find a bimodal
distribution of heliocentric (line-of-sight) radial velocities,
$v_{\rm los}$.  The distribution of radial velocities has two significant
peaks falling in the range $10<R_{\rm GC}<11$ kpc at $v_{\rm los}=-4$ km
s$^{-1}$ and $v_{\rm los}=+27$ km s$^{-1}$.  Since the field is directed
towards the anti-Centre, the line-of-sight velocities $v_{\rm los}$ serve
as a good proxy for the galactocentric radial velocities, $v_{R}$.
For the anti-Centre, we have
\begin{equation}
\label{ }
v_{\rm GSR} = v_{\rm los}-\text{U} \approx v_{R}
\end{equation}
By correcting for the radial Solar motion, which has an inward motion
of $10\lesssim \text{U} \lesssim14$ km s$^{-1}$, the distribution
becomes symmetric about $v_{GSR}=0$ km s$^{-1}$ with peaks at roughly
$\pm15$ km s$^{-1}$.

\citet{Liu2012a} suggest that this feature is due to a resonant
interaction with the Milky Way's bar, since their locations are
consistent with the position of the Outer Lindblad Resonance.  The
bimodal distribution in our simulation occurs outside the bar's OLR
($\sim$8.5 kpc).  Since our disc is kinematically hot
($\sigma_{v_{R}}\approx65$ km s$^{-1}$ at $R=9$ kpc), the difference
between the two peaks in the velocity distribution is larger than the
observations.  We have however provided a possible mechanism for such
a feature to arise.  If the resonance is populated in a hot disk, then
it is likely to be still more important in a colder disk. In hotter
disks, the larger velocity dispersions of stars means they are less
likely to become and remain trapped in narrow resonances.

Finally, in the right plot of Figure \ref{fig:vrDenseCut}, we show the
time-averaged imprint of the 3:-2 orbits in $R-v_{R}$ space.  The
vertical and horizontal extent of the distribution is a measure of the
epicycle energy associated with the orbits and the density can be
thought of as a measure of the time spent by the particles in these
regions.  The particles spend little time at pericentre with the
majority of the orbit spent at large $R$.  Kinematic structures due to
the resonant orbits are therefore more persistent if they occur in the
outskirts of the disc.

\section{Discussion}\label{sec:Discuss}

To our knowledge, there has been little investigation of resonant
orbits in the outskirts of barred N-body discs.  Resonant clumping
therefore seems to be a previously unknown phenomenon.  For this
reason, we discuss the mechanism that cause the periodic orbits to
become clumped in phase angle ($\theta_{R}$) and to remain so for long
periods in the evolution in the disc.  In the following, we address
two questions: (1) How do the periodic orbits become clumped in the
first place? and (2) Why do they remain clumped when phase mixing
should work to undo the clumping?

\subsection{Resonant Clumping: Origins}
\begin{figure*}[t]
\begin{center}
\includegraphics[width=\textwidth]{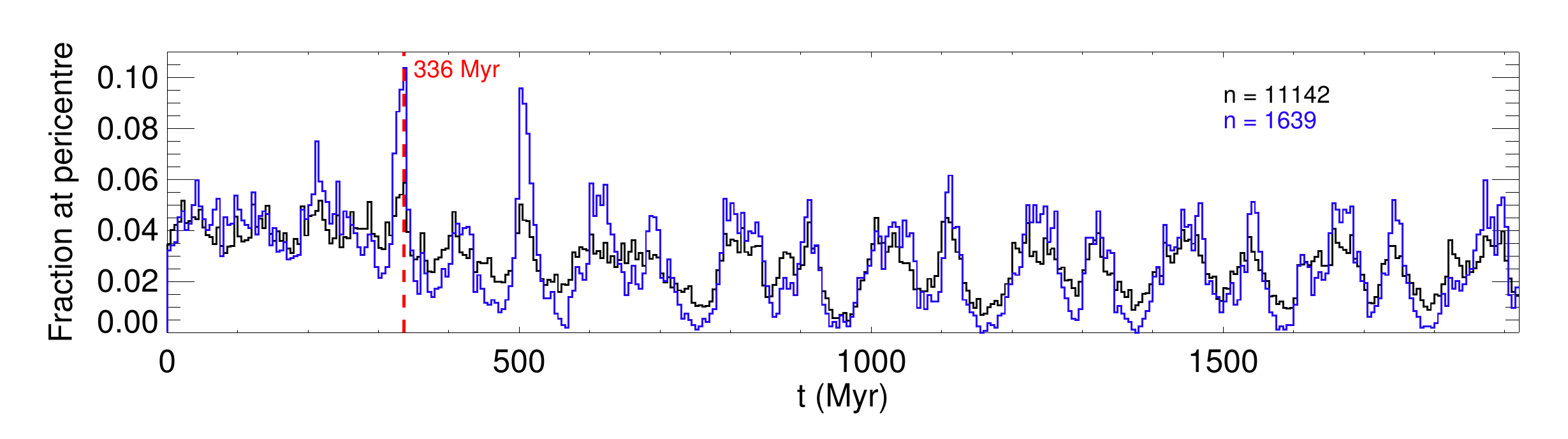}
\includegraphics[width=\textwidth]{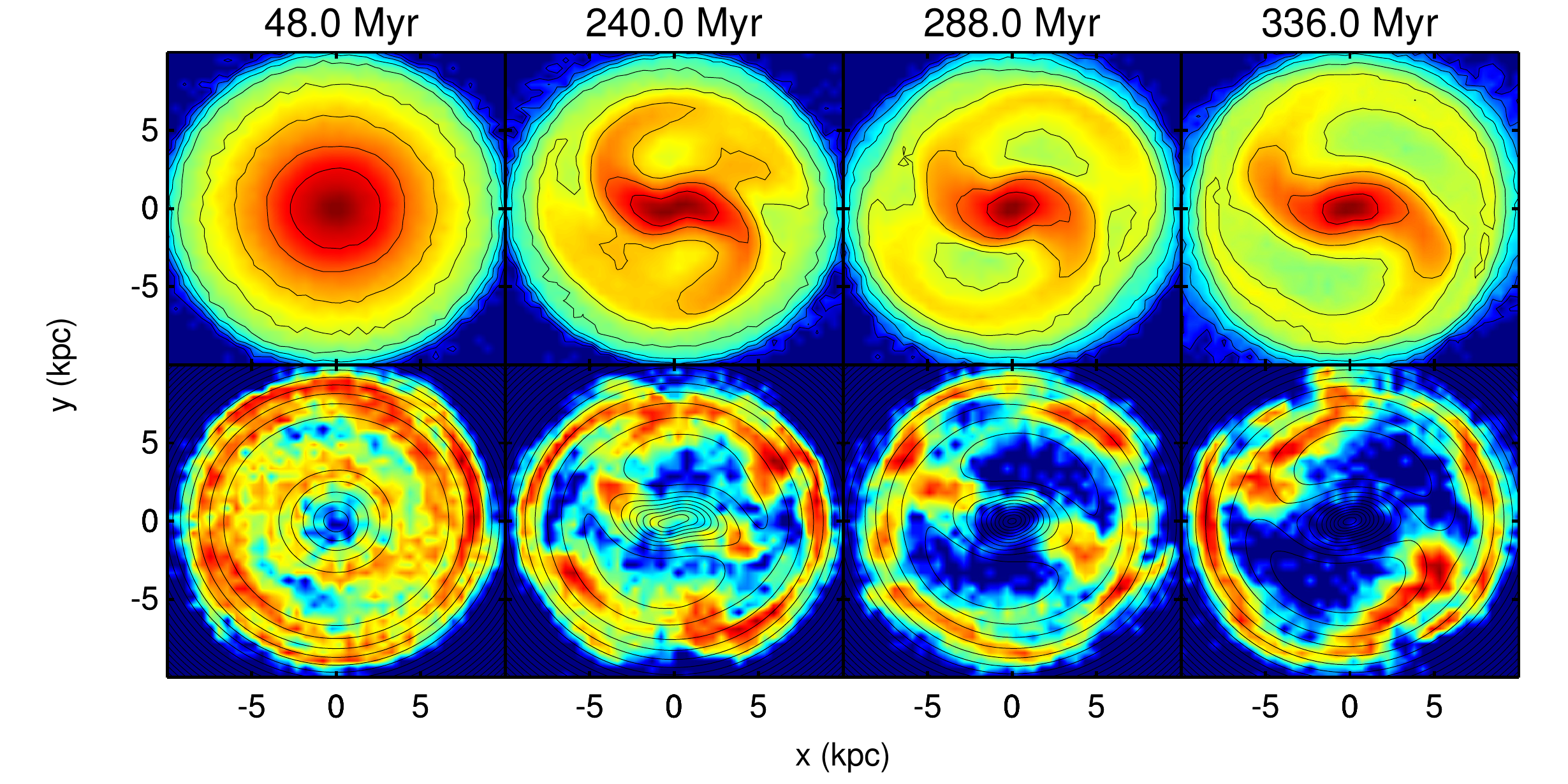}
\caption{Clumping occurs early on in the evolution of the disc. In the
  top panel, we show the distribution of pericentre times for
  particles extracted from segment two (black) and particles common to
  segments two and eight (blue). The surface density of the disc
  during the formation of the bar is shown in the top row of the
  bottom panel (contours of equal surface density). The resonant
  particles exhibit overdensities at the Lagrange points at either end
  of the bar (contours of equal effective potential, as shown in the
  final row).}\label{fig:ResOrig}
\end{center}
\end{figure*}

To investigate the first point, we apply the phase space distance
method described in Section \ref{sec:PDM} over many epochs during the
evolution of the disc.  Specifically, we use eight $\sim$1 Gyr time
segments, from $\sim$0.5 Gyr to $\sim$1.5 Gyr, $\sim$1.0 Gyr to
$\sim$2.0 Gyr, $\sim$1.5 Gyr to $\sim$2.5 Gyr all the way up to
$\sim$4.0 to $\sim$5.0 Gyr, which was the segment we analysed in the
previous sections of the paper.  We avoid using data from the epoch of
bar formation (0-500 Myr), since angular momentum is violently
redistributed throughout the disc.  The resulting variations in the
guiding radii and rotational speeds make the normalisations on the
phase space distance unreliable.  In any case, we are able to extract
resonant orbits from all eight segments.  The number of resonant
orbits increases with time and begins to flatten out after segment
five (i.e, from $\sim$2.5 Gyr onwards).  We also find, by analysing
the azimuthal and radial frequencies, that the 3:-2 orbit family is
the dominant family even from early on.

For our current purposes, we are interested in particles that are
captured into resonance early on and that remain in resonance for a
long time.  To select such a long-lived resonant sample, we take
particles found during segment two ($\sim$1 Gyr to $\sim$2 Gyr) and
compare them to the particles found during segment eight ($\sim$4 Gyr
to $\sim$5 Gyr).  Particles that are common to both samples have
remained in resonance for at least $\sim$4 Gyr.

Due to the changes in angular momentum (and the corresponding changes
in $R_{\rm g}$ and $\Omega$), especially during the formation of the bar,
a reliable measure of the phase angle $\theta_{R}$ (i.e., the clumped
angle) is unavailable to us.  For this reason, we continue to use the
times at which the particles reach pericentre as an indicator of
whether clumping in the phase angle is present or not.  While this is
a robust indication of the presence of clumping, we are unable to say
if the particles become clumped in the time period between apocentre
and pericentre.  In Figure \ref{fig:ResOrig} (top panel), we show the
distribution of pericentre times for the 3:-2 particles extracted from
segment two ($\sim$1 Gyr to $\sim$2 Gyr; black) and the 3:-2 particles
that are common to segment two and segment eight (the long-lived
resonant particles; blue).  The distributions are shown from $t=0$ Gyr
to $t=\sim$2 Gyr and indicate that the particles become clumped during
the formation of the bar (between 200 Myr and 500 Myr).

In the bottom panel of Figure \ref{fig:ResOrig}, we show the surface
density of all particles in the disc (top row) during the formation of
the bar with contours of equal density.  The bar in this case forms
very quickly (within 300 Myr) and its arrival induces strong spiral
structures.  We also show the surface density for our sample of 3:-2
particles from segment two in the bottom rows, this time with contours
of equal effective potential.  The outline of the resonant orbits
exhibits a striking resemblance to the large ringed-shaped features
seen in external barred galaxies (the so-called `R' rings of
\citet{Buta1986a}). It is also clear that overdensities appear close
to the Lagrange points at either end of the bar (i.e., L$_{1}$ and
L$_{2}$).  Given that (a) the periodic orbits are clumped in
$\theta_{R}$ as the bar is forming, (b) the bar induces strong
spirals, (c) the resonant particles appear to follow the pattern of
rings in barred galaxies and (d) overdensities appear close to the
Lagrange points, we naturally suspect that the clumping is related to
the mechanism that generates these type of structures -- namely,
passage through the unstable Lagrange points and their associated
manifolds.

\begin{figure}[t]
\begin{center}
\includegraphics[width=0.5\textwidth]{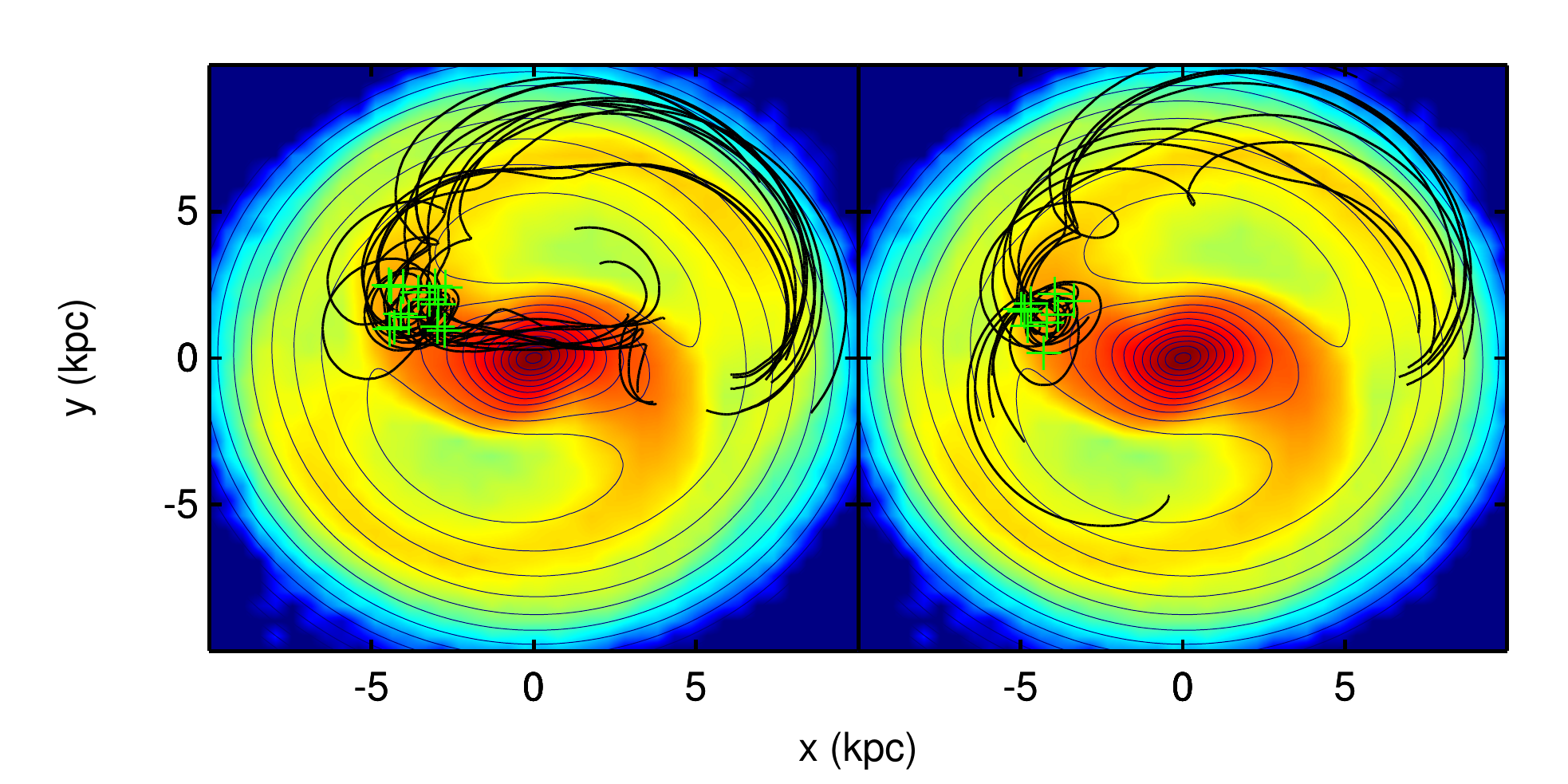}
\caption{The surface density of all particles at $\sim$200 Myr with
  contours of equal effective potential $\Phi_{\text{\rm eff}}$. Left
  panel: Trajectories of periodic orbits that approach L$_{1}$ from
  the bar. Right panel: Trajectories of periodic orbits that approach
  from the outer disc. The orbits are plotted from $\sim$100 to 450
  Myr. The green crosses \textbf{indicate} the positions at $\sim$200 Myr. The
  particles move in a clockwise direction (in the frame corotating
  with the bar).}\label{fig:ManifoldOrbits}
\end{center}
\end{figure}

The mechanics of bar induced spirals and large scale rings in barred
galaxies has been studied in detail
\citep[cf.][]{Gomez2004,Romero-Gomez2006,Romero-Gomez2007,Athanassoula2009a,Athanassoula2009b}.
The first main point relevant to our work is that L$_{1}$ (and
L$_{2}$) are host to the Lyapunov orbits (roughly elliptical orbits
that corotate with $L_{1/2}$).  Since the region is unstable, the
periodic Lyapunov orbits can't indefinitely trap other regular orbits.
Orbits that visit this region tend to leave it on a timescale
proportional to their energy, i.e., lower energy orbits escape the
region faster than orbits with a higher energy.  The second main point
is that the trajectories through which the orbits (which may have been
temporarily trapped) can leave this region are determined by the
unstable manifolds -- whereas the stable manifolds describe the
trajectories through which orbits approach this region.  As an
example, we plot some orbits that follow these manifolds in Figure
\ref{fig:ManifoldOrbits}.  Note that not all of the resonant orbits
approach the Lagrange points.  From the total sample of resonant
particles, we estimate that about 75\% approach the Lagrange points.
For these, the average radius at $t=0$ Myr is $R_{0}\approx4$ kpc with
$2<R_{0}<7$ kpc.  In the left (right) panel, we show orbits that come
from the inner (outer) parts of the disc and in the right.  At
$t\approx150$ Myr, the majority of particles have $R<4$ kpc indicating
passage along the inner stable manifold (i.e., along the bar).  The
green crosses indicate the positions corresponding to the time at
which the surface density and equipotential contours are plotted.  The
orbits can be compared with those shown in Figure 4 of
\citet{Athanassoula2012}.

The third, and possibly most important, point in this context is that
the L$_{1}$ and L$_{2}$ points can be stabilised by concentrations of
matter \citep[see section 5 of][]{Athanassoula2009b}.  This generates
a local minimum in the potential, a stable equilibrium point
surrounded by two unstable saddle points (each with their own
manifolds).  This stable minimum can trap particles and keep them in
the vicinity for a long period of time.  As more particles gather
here, so the minimum in the potential deepens, and the overdensity
grows. This seems a plausible mechanism for the initial clumping.

We have shown that at the epoch of bar formation, the phase angles are
clumped, as they appear to reach pericentre at the same time (Figure
\ref{fig:ResOrig}, top panel).  We have also shown that the orbits
pass close to the unstable Lagrange points on either end of the bar
(Figure \ref{fig:ResOrig}, bottom panel) and as they exit they follow
the unstable manifolds (Figure \ref{fig:ManifoldOrbits}) and exhibit a
structure similar to the manifold-driven spirals.  Since we can only
probe the clumping at apo/pericentre, we can't demonstrate whether the
clumping in the phase angle occurs before, during or after the
particles move through the Lagrange points.  So, while we can only
speculate as to the mechanism by which the phase angles become
coherent -- due to passage through the unstable equilibrium points -- we
can conclusively say that the initial clumping occurs during the
formation of the bar.  A rigorous investigation of the mechanism that
causes the clumping of the phase angles requires an experiment in
which the phase angles can be measured at all points along the orbit.

\subsection{Resonant Clumping: Persistence}

\begin{figure*}[t]
\begin{center}
\includegraphics[width=\textwidth]{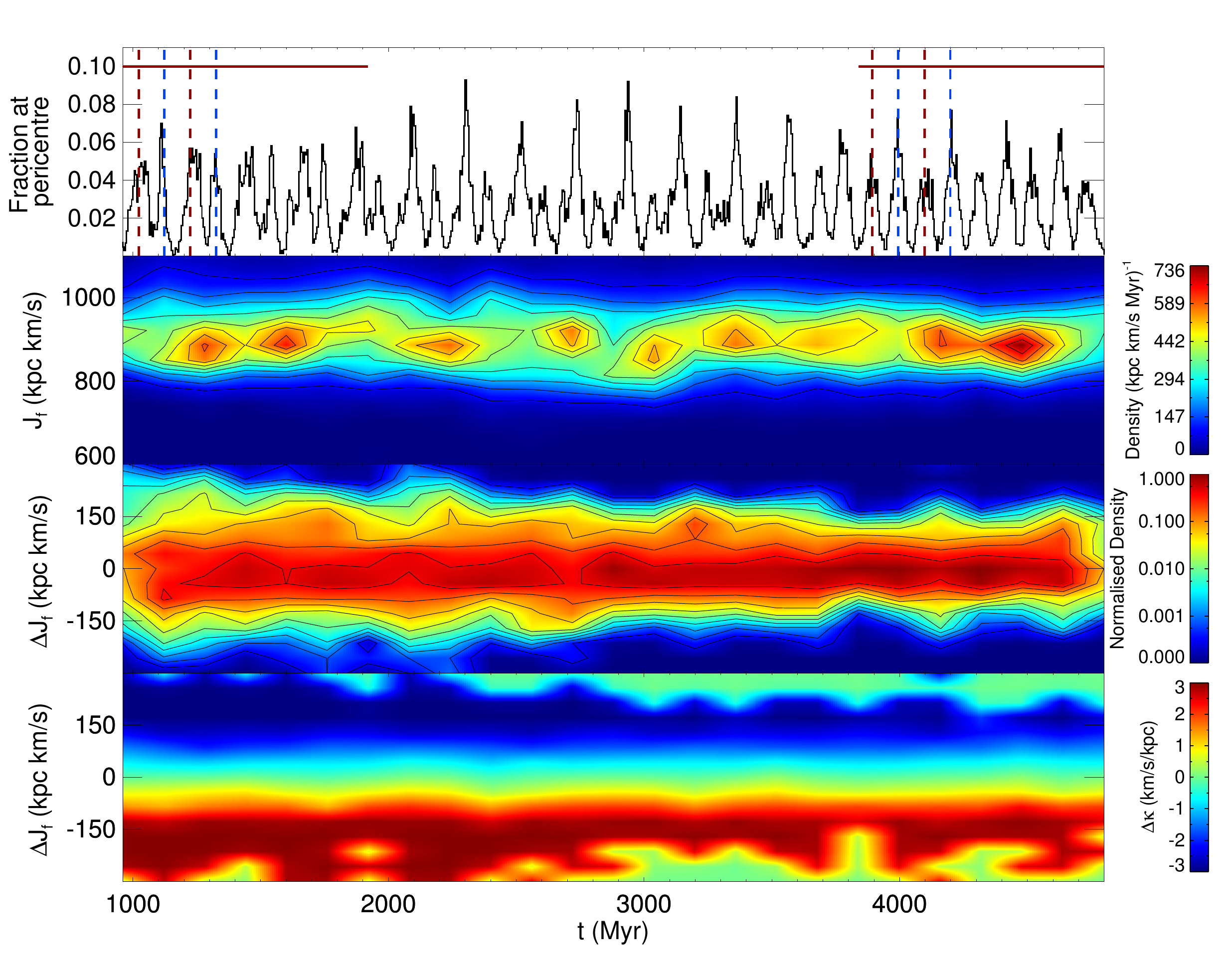}
\caption{Phase mixing is inefficient, as the clumping in the phase
  angle $\theta_{R}$ persists for long periods. The pericentre times
  are shown for the resonant particles common to segments two and
  eight (highlighted with red on the horizontal axis). The overall
  distribution of fast actions ($J_{\rm f}$ second row) remains
  unchanged over the $\sim$4 Gyr shown here. Individual particles do
  however experience changes in their respective actions shown by the
  distribution of $\triangle J_{\rm f}$ in the third panel. Changes in the
  fast action are accompanied by changes in the corresponding
  frequency $\kappa$ (where
  $\dot{\theta}_{f}=\dot{\theta}_{R}=\kappa$). }\label{fig:ClumpPersist}
\end{center}
\end{figure*}

Since the clumping in the phase angle $\theta_{R}$ occurs very early
on, it is natural to expect that, as the disc evolves, the phase angle
becomes mixed and the clumping should become less prominent. This is
not the case however.  For our sample of long-lived resonant particles
(from the blue histogram, top panel, Figure \ref{fig:ResOrig}), we see
that the clumping in their phase angles persists for nearly 4 Gyr,
apparently becoming more coherent as time progresses.  The top plot in
Figure \ref{fig:ClumpPersist} shows the distribution of pericentre
times for this sample.  The clumping occurs on timescales comparable
with the radial period of the epicycle for these particles (with
$\kappa\approx30$ km/s/kpc, $T_{R}\approx205$ Myr) and remains so for
almost the whole duration of the simulation.  This is clear evidence
that something is counteracting the effects of phase mixing.

Since the phenomenon of resonant clumping necessarily involves
transient overdensities, we suggest that changes in the actions
$J_{R}$ and $J_{\phi}$ are working against phase mixing and keeping
them locked in phase angle.  Below we set out how we measure the
actions.  Note we do this only for the particles caught in segments 2
\& 8 that are deemed to be in a 3:-2 resonance -- i.e., those from the
top panel of Figure \ref{fig:ClumpPersist}.

We compute the actions ($J$) by numerically integrating over the
N-body trajectories.  We measure the radial action $J_{R}$ as
\begin{equation}
\label{eqn:JR}
J_{R}=\dfrac{1}{\pi}\int_{R_{\rm min}}^{R_{\rm max}}v_{R}dR
\end{equation}
where we have integrated over half of the epicycle.  The next
measurement is taken over the second half of the epicycle.  To follow
the evolution of the actions, we assign a time to each measurement.
If we begin the measurement at $t(R_{1})$ and end it at $t(R_{2})$
with the duration $\triangle t=t(R_{2})-t(R_{1})$, then we say that
orbit has an action $J_{R}$ at time $t=t(R_{1})+\triangle t/2$.  It is
also convenient to measure the epicyclic frequency $\kappa$ as
\begin{equation}
\label{ }
\kappa=\dfrac{\pi}{t(R_{2})-t(R_{1})}. 
\end{equation}
We measure the azimuthal action as 
\begin{equation}
\label{ }
J_{\phi}=\dfrac{1}{\pi}\int_{\phi_{1}}^{\phi_{2}}L_{z}d\phi
\end{equation} 
where $\phi_{1}$ and $\phi_{2}$ are the azimuths at which pericentre
or apocentre is reached, i.e., the azimuths at $t(R_{1})$ and
$t(R_{2})$ from above.  In a similar way to our measurement of
$\kappa$, we reckon $\Omega$ as
\begin{equation}
\label{ }
\Omega=\dfrac{\phi_{2}-\phi_{1}}{t(R_{2})-t(R_{1})}.
\end{equation}
The fast action $J_{\rm f}$ is just a linear combination of $J_{R}$
and $J_{\phi}$ \citep[see][for details]{Collett1997}, so that
\begin{equation}
\label{eqn:fast}
J_{\rm f}=J_{R}+\dfrac{|l|}{m}J_{\phi}
\end{equation}
where $m$ \& $l$ are the usual integers describing the closed orbit.
$J_{\rm f}$ describes the motions of particles along the orbit pattern
while the slow action, $J_{\rm s}\approx J_{\phi}$, describes the
precession of the apsides of the orbit pattern (the 3:-2 pattern).

Figure \ref{fig:ClumpPersist} (second panel) shows the evolution of
the fast actions for the sample of resonant particles extracted from
time segments 2 \& 8.  The combination of slightly increasing radial
actions and slightly decreasing azimuthal actions (not shown) leads to
a distribution of fast actions that barely changes.  We have also made
an estimate of $\triangle J_{\rm f}$ and $\triangle\kappa$, by taking
one measurement of $J_{\rm f}$ or $\kappa$ subtracted from the next.
We remark that this is not the time derivative of $J_{\rm f}$, but the
change per half epicycle and gives us an idea of the trends in the
actions and frequencies.  Figure \ref{fig:ClumpPersist} (third panel)
shows that although the overall distribution of $J_{\rm f}$ does
remain largely constant, there \textit{are} changes in the actions of
individual particles.  The changes are most pronounced at earlier
times, but persist until the later stages of the disc's evolution.
Figure \ref{fig:ClumpPersist} (fourth panel) then shows that the
changes in $J_{\rm f}$ ($\triangle J_{f}$) are accompanied by changes
in $\kappa$ ($\triangle\kappa$).  This is exactly the same as Figure
\ref{fig:ClumpPersist} (third panel) except instead of colour
representing density, it now represents the average change in $\kappa$
-- blue represents a decrease in $\kappa$ and red represents an
increase.  If, for a particular particle, the fast action has
increased, then this is accompanied by a decrease in the corresponding
frequency - which is $\kappa$ in this case.  The opposite occurs for a
decrease in $J_{\rm f}$. The distributions of pericentre times shows
that mixing in the phase angle is not occurring, so that the changes
in the frequency $\kappa$ work against the phase mixing.

\section{Summary \& Conclusions}

With the emergence of N-body simulations as the workhorse in the study
of galaxy dynamics and evolution, it is important to have the tools
available to extract as much information as possible from them. The
potential in galaxy discs is inherently time-dependent, driven by
evolving bars, rings and spiral patterns. This then prevents a steady
state description of the disc using a time-independent Hamiltonian.
Frequency analyses also encounter issues in highly dynamic, unsteady
and evolving systems.  When low frequency orbits occur in varying
potentials, it can be hard to separate them from induced radial modes
\citep[e.g.,][]{Ceverino2007}.  In discs with significant angular
momentum exchange, it can also be difficult to measure the
characteristic frequencies of the orbits ($\kappa$ and $\Omega$).

Resonant orbits play a special role in such systems.  We have
developed a method to make rapid and automated searches for closed
orbits in N-body simulations. This algorithm utilises the fact that
orbits resonating with a bar or spiral pattern in a rotating frame
return periodically to previously inhabited spot in phase space.  By
defining a metric that measures distance travelled in phase space, we
have extracted samples of closed orbits from the simulation of
\citet{Shen2010}.  The sample can be tuned by use of an arbitrary
phase space distance cutoff point.  The choice of cutoff depends on
how ``clean" a sample of closed orbits is required. This essentially
separates an N-body simulation into distinct components - the,
possibly many, resonant families and the background disc population.
The method has its greatest utility in complex dynamical systems such
as a Galactic bar where perturbation theory and frequency analysis may
have difficulty (see e.g., Molloy et al 2014b).

As a worked application, we use the method to dissect the N-body
simulation of \citet{Shen2010}. This starts as an unstable disc that
rapidly forms a massive bar, which evolves through buckling
instabilities to give a bulge that is an excellent match to the data
in the central parts of the Milky Way. However, a drawback to the
simulation is that the outer disc is kinematically hotter than is the
case for the Milky Way, which prevents a direct comparison with
data. The size of the bar and the absence of both gas and a live halo
mean that angular momentum exchange in the disc is maximised.  These
two aspects of the simulation mean that perturbation theory and a
frequency analysis would rapidly encounter their respective
limitations.

In the outer disc, we extracted a sample of resonant orbits.  By
combining the orbital properties of our sample with Fourier
spectrograms, we have demonstrated the source of the resonance as the
central bar itself (Figure \ref{fig:PSpecOverplot}).  In this
simulation, the major populations of resonant orbits are the 3:-2 and
1:-1 families.  By measuring the epochs at which closed orbits reach
their turning points, we see that they move in a coordinated fashion.
The particles that populate these resonant orbits all conspire to
reach their pericenters and apocentres at roughly the same time
(Figure \ref{fig:TPointsxy}).  This leads to the phenomenon of {\it
  resonant clumping} in the disc. Overdensities are produced each time
a pericenter or apocentre is reached. These overdensities can make a
significant contribution to the disc potential and their transient
nature prohibits a description of the disc by means of a
time-independent Hamiltonian.

The family of 3:-2 orbits exist in two distinct groups, whose
epicycles are out of phase by $\pi$.  This means that as one group
reaches apocentre, the other reaches pericentre.  As the particles
complete their respective epicyclic motions, they overlap in
configuration space with one group moving inwards, and the other
moving outwards.  This characteristic of their collective motion
generates a bimodal distribution of galactocentric radial velocities
that occurs at distinct angles with respect to the bar.  For the
family of 3:-2 orbits, this occurs at angles of between
20-40\textdegree\ which is in the range measured for the viewing angle
of the Galactic bar.  While we refrain from making a direct comparison
with the Milky Way due to the hotness of the simulated disc, we have
provided a mechanism which may explain the bimodal distribution of
$v_{R}$ observed by \citet{Liu2012a} towards the anti-Centre.

Each group can also be divided into subgroups depending on the azimuth
at which they reach their pericenter or apocentre.  For example, if
one subgroup reaches their pericentre on the negative $x$-axis (i.e.,
at one end of the bar), another subgroup appears on the positive
$x$-axis.  This means that, as one overdensity is produced, so another
is produced on the opposite side of the disc.  This maintains a
bisymmetric pattern and prevents the disc from becoming lopsided.

The phenomenon of resonant clumping has not received much attention
hitherto.  For this reason, we have endeavored to uncover how it comes
about, and why it persists in the disc.  The global morphology in the
density distribution of the resonant particles along with their
proximity to the unstable Lagrange points (and their approach and
departure) suggests there is a strong link between the initial
clumping mechanism and that which brings about bar-induced spirals --
namely, passage through the Lagrange points.  Unfortunately, we have
been unable to conclusively show how the phase angles become clumped
since a reliable measure of $\theta_{R}$ along all points of the orbit
is unavailable to us.  That the clumping occurs during the formation
of the bar is, however, beyond doubt. Since it is likely associated
with bar-induced spirals, we suggest that clumping may be responsible
for the incomplete, or partial, rings in barred galaxies (the
``pseudorings'' or R$^{\prime}$ rings of \citet{Buta1986a}).  It has been
shown that these rings can be generated by stars ejected along the
unstable manifolds \cite[e.g.,][]{Athanassoula2012}.  If the rate of
ejection from the Lagrange points is not constant (due to, say,
clumping in the phase angle), then we expect incomplete rings to form.
The clumping maintains itself by counteracting phase mixing.  The
transient overdensities provide a force that alters the orbit's
actions and corresponding frequencies.  Since we have used very little
prior information in extracting these resonant orbits (only the choice
of normalisation), we have preserved the time dependent transient
nature of processes in the disc.

The release of the Gaia \citep{Perryman2001} and LAMOST
\citep{Deng2014} datasets will herald a dramatic advance in our
knowledge of the kinematic landscape, not only in the Solar vicinity
but also to the very edges of the disc.  For this reason, it is vital
that we have at our disposal the tools required to analyse fully the
output from increasingly sophisticated N-body simulations.  Deviations
from axisymmetry and the resulting resonances must be well understood
if we are to comprehend how the Milky Way has evolved.  Improving our
knowledge of the kinematic structures will help us understand better
the importance of secular mechanisms at work in the disc.  Decomposing
N-body discs into their major orbital components gives us the
opportunity to begin to build up analytic models of complex systems by
superposing many distinct distribution functions.  The method
presented here for dissecting simulations should serve as a
complementary tool to more established techniques and has the
potential to fill in the gaps in our knowledge of galaxy evolution
that have yet to be illuminated.

\acknowledgements We are deeply indebted to Marcel Zemp \& Zhao-Yu Li
for many helpful discussions and support during this project.  We also
thank the referee for insightful comments which greatly improved the
paper.  The authors acknowledge financial support from the CAS One
Hundred Talent Fund and NSFC Grants 11173002, 11333003, 11322326 and
11073037.  This work was also supported by the following grants: the
Gaia Research for European Astronomy Training (GREAT-ITN) Marie Curie
network, funded through the European Union Seventh Framework Programme
(FP7/2007-2013) under grant agreement no 264895; the Strategic
Priority Research Program ``The Emergence of Cosmological Structures''
of the Chinese Academy of Sciences, Grant No. XDB09000000; and the
National Key Basic Research Program of China 2014CB845700. This work made use of the super-computing facilities at Shanghai Astronomical Observatory.

\bibliography{Fourth_Draft_Refs}

\end{document}